\documentclass[twocolumn,tighten]{aastex62}
\usepackage{graphicx}
\graphicspath{{./figures/}{./}}

\usepackage{listings}
\usepackage{enumitem}
\usepackage{lineno}
\usepackage{comment}
\input{hyperlink-year-only-natbib-patch}

\usepackage{soul} 
\usepackage{amsmath}
\usepackage{amssymb}
\usepackage{xspace}
\usepackage{xifthen}

\newcommand{\ie}{i.e.\xspace}
\newcommand{\eg}{e.g.\xspace}

\newcommand{\FIXME}[1]{{}}
\newcommand{\CHECK}[1]{{#1}}
\newcommand{\COMMENT}[3]{{}}

\mathchardef\mhyphen="2D
\newcommand{\vect}[1]{\boldsymbol{#1}}
\newcommand{\roughly}{\ensuremath{ {\sim}\,} }

\newlength{\dhatheight}

\newcommand{\code}[1]{\texttt{#1}\xspace}
\newcommand{\dd}{\ensuremath{\rm d}}

\newcommand{\unit}[1]{\ensuremath{\mathrm{\,#1}}\xspace}
\newcommand{\Gyr}{\unit{Gyr}}

\newcommand{\amin}{\unit{arcmin}}
\newcommand{\asec}{\unit{arcsec}}

\newcommand{\pc}{\unit{pc}}
\newcommand{\kpc}{\unit{kpc}}

\newcommand{\Msun}{\unit{M_\odot}}

\newcommand{\magn}{\unit{mag}}
\newcommand{\mmag}{\unit{mmag}}

\newcommand{\secref}[1]{Section~\ref{sec:#1}}
\newcommand{\appref}[1]{Appendix~\ref{app:#1}}
\newcommand{\tabref}[1]{Table~\ref{tab:#1}}
\newcommand{\figref}[1]{Figure~\ref{fig:#1}}
\newcommand{\eqnref}[1]{Equation~\eqref{eqn:#1}}

\newcommand{\secrefs}[2]{Sections~\ref{sec:#1} and \ref{sec:#2}}
\newcommand{\tabrefs}[2]{Tables~\ref{tab:#1} and \ref{tab:#2}}
\newcommand{\figrefs}[2]{Figures~\ref{fig:#1} and \ref{fig:#2}}

\newcommand{\bandvar}[2][]{%
  \ifthenelse{\isempty{#1}}{\var{#2}}{\var{#2\_#1}}%
}

\newcommand{\modulus}{\ensuremath{m - M}\xspace}

\newcommand{\ra}{{\ensuremath{\alpha_{2000}}}\xspace}
\newcommand{\dec}{{\ensuremath{\delta_{2000}}}\xspace}
\newcommand{\age}{{\ensuremath{\tau}}\xspace}
\newcommand{\metal}{{\ensuremath{Z}}\xspace}

\newcommand{\ellip}{\ensuremath{\epsilon}\xspace}
\newcommand{\PA}{\ensuremath{\mathrm{P.A.}}\xspace}

\newcommand{\ngmix}{\code{ngmix}}

\newcommand{\SExtractor}{\code{Source{\allowbreak}Extractor}}
\newcommand{\sextractor}{\SExtractor}

\newcommand{\HEALPix}{\code{HEALPix}}
\newcommand{\healpix}{\HEALPix}
\newcommand{\nside}{\code{nside}}

\newcommand{\mangle}{\code{mangle}}

\newcommand{\ugali}{\code{ugali}}
\newcommand{\simple}{\code{simple}}
\newcommand{\var}[1]{\ensuremath{\texttt{\MakeUppercase{#1}}}\xspace}
\newcommand{\Gaia}{{\it Gaia}\xspace}

\newcommand{\TS}{\ensuremath{\mathrm{TS}}\xspace}
\newcommand{\SIG}{\ensuremath{\mathrm{SIG}}\xspace}
\newcommand{\Pdet}{\ensuremath{P_{\rm det}}\xspace}

\providecommand\physrep{\ref@jnl{Phys.~Rep.}}%
\providecommand\apjs{\ref@jnl{ApJS}}%
\providecommand{\jcap}{\ref@jnl{JCAP}}%

\newcommand{\Nstars}{\ensuremath{N(g < 22)}\xspace}

\newcommand{\bp}[1]{\ensuremath{#1_{\rm P1}}\xspace}
\newcommand{\gp}{\bp{g}}
\newcommand{\rp}{\bp{r}}
\newcommand{\ip}{\bp{i}}
\newcommand{\zp}{\bp{z}}
\newcommand{\yp}{\bp{y}}

\newcommand{\cetII}{{DES\,J0117\allowbreak$-$1725}\xspace}

\defcitealias{DR1:2018}{DES Collaboration (2018)}

\newcommand{\KB}[1]{\COMMENT{brown}{KB}{#1}}

\makeatletter
\providecommand*{\input@path}{}
\g@addto@macro\input@path{{./tables/}}
\makeatother

\usepackage{hyperref}

\shorttitle{Milky Way Satellite Census -- I.}
\shortauthors{Drlica-Wagner \& Bechtol et al.}

\begin{document}

\title{Milky Way Satellite Census. I. The Observational Selection Function for Milky Way Satellites in DES Y3 and Pan-STARRS DR1}


\author[0000-0001-8251-933X]{A.~Drlica-Wagner}
\affiliation{Fermi National Accelerator Laboratory, P. O. Box 500, Batavia, IL 60510, USA}
\affiliation{Kavli Institute for Cosmological Physics, University of Chicago, Chicago, IL 60637, USA}
\affiliation{Department of Astronomy and Astrophysics, University of Chicago, Chicago, IL 60637, USA}
\author[0000-0001-8156-0429]{K.~Bechtol}
\affiliation{Physics Department, 2320 Chamberlin Hall, University of Wisconsin-Madison, 1150 University Avenue Madison, WI  53706-1390}
\affiliation{LSST, 933 North Cherry Avenue, Tucson, AZ 85721, USA}
\author[0000-0003-3519-4004]{S.~Mau}
\affiliation{Kavli Institute for Cosmological Physics, University of Chicago, Chicago, IL 60637, USA}
\affiliation{Department of Astronomy and Astrophysics, University of Chicago, Chicago, IL 60637, USA}
\author[0000-0001-5435-7820]{M.~McNanna}
\affiliation{Physics Department, 2320 Chamberlin Hall, University of Wisconsin-Madison, 1150 University Avenue Madison, WI  53706-1390}
\author[0000-0002-1182-3825]{E.~O.~Nadler}
\affiliation{Department of Physics, Stanford University, 382 Via Pueblo Mall, Stanford, CA 94305, USA}
\affiliation{Kavli Institute for Particle Astrophysics \& Cosmology, P. O. Box 2450, Stanford University, Stanford, CA 94305, USA}
\author[0000-0002-6021-8760]{A.~B.~Pace}
\affiliation{Department of Physics, Carnegie Mellon University, Pittsburgh, Pennsylvania 15312, USA}
\affiliation{George P. and Cynthia Woods Mitchell Institute for Fundamental Physics and Astronomy, and Department of Physics and Astronomy, Texas A\&M University, College Station, TX 77843,  USA}
\author[0000-0002-9110-6163]{T.~S.~Li}
\altaffiliation{NHFP Einstein Fellow}
\affiliation{Observatories of the Carnegie Institution for Science, 813 Santa Barbara St., Pasadena, CA 91101, USA}
\affiliation{Department of Astrophysical Sciences, Princeton University, Peyton Hall, Princeton, NJ 08544, USA}
\affiliation{Fermi National Accelerator Laboratory, P. O. Box 500, Batavia, IL 60510, USA}
\affiliation{Kavli Institute for Cosmological Physics, University of Chicago, Chicago, IL 60637, USA}
\author[0000-0001-9186-6042]{A.~Pieres}
\affiliation{Laborat\'orio Interinstitucional de e-Astronomia - LIneA, Rua Gal. Jos\'e Cristino 77, Rio de Janeiro, RJ - 20921-400, Brazil}
\affiliation{Observat\'orio Nacional, Rua Gal. Jos\'e Cristino 77, Rio de Janeiro, RJ - 20921-400, Brazil}
\author[0000-0002-1666-6275]{E.~Rozo}
\affiliation{Department of Physics, University of Arizona, Tucson, AZ 85721, USA}
\author{J.~D.~Simon}
\affiliation{Observatories of the Carnegie Institution for Science, 813 Santa Barbara St., Pasadena, CA 91101, USA}
\author[0000-0002-7123-8943]{A.~R.~Walker}
\affiliation{Cerro Tololo Inter-American Observatory, National Optical Astronomy Observatory, Casilla 603, La Serena, Chile}
\author[0000-0003-2229-011X]{R.~H.~Wechsler}
\affiliation{Department of Physics, Stanford University, 382 Via Pueblo Mall, Stanford, CA 94305, USA}
\affiliation{Kavli Institute for Particle Astrophysics \& Cosmology, P. O. Box 2450, Stanford University, Stanford, CA 94305, USA}
\affiliation{SLAC National Accelerator Laboratory, Menlo Park, CA 94025, USA}
\author{T.~M.~C.~Abbott}
\affiliation{Cerro Tololo Inter-American Observatory, National Optical Astronomy Observatory, Casilla 603, La Serena, Chile}
\author[0000-0002-7069-7857]{S.~Allam}
\affiliation{Fermi National Accelerator Laboratory, P. O. Box 500, Batavia, IL 60510, USA}
\author[0000-0002-0609-3987]{J.~Annis}
\affiliation{Fermi National Accelerator Laboratory, P. O. Box 500, Batavia, IL 60510, USA}
\author{E.~Bertin}
\affiliation{CNRS, UMR 7095, Institut d'Astrophysique de Paris, F-75014, Paris, France}
\affiliation{Sorbonne Universit\'es, UPMC Univ Paris 06, UMR 7095, Institut d'Astrophysique de Paris, F-75014, Paris, France}
\author[0000-0002-8458-5047]{D.~Brooks}
\affiliation{Department of Physics \& Astronomy, University College London, Gower Street, London, WC1E 6BT, UK}
\author{D.~L.~Burke}
\affiliation{SLAC National Accelerator Laboratory, Menlo Park, CA 94025, USA}
\affiliation{Kavli Institute for Particle Astrophysics \& Cosmology, P. O. Box 2450, Stanford University, Stanford, CA 94305, USA}
\author[0000-0003-3044-5150]{A.~Carnero~Rosell}
\affiliation{Centro de Investigaciones Energ\'eticas, Medioambientales y Tecnol\'ogicas (CIEMAT), Madrid, Spain}
\affiliation{Laborat\'orio Interinstitucional de e-Astronomia - LIneA, Rua Gal. Jos\'e Cristino 77, Rio de Janeiro, RJ - 20921-400, Brazil}
\author[0000-0002-4802-3194]{M.~Carrasco~Kind}
\affiliation{National Center for Supercomputing Applications, 1205 West Clark St., Urbana, IL 61801, USA}
\affiliation{Department of Astronomy, University of Illinois at Urbana-Champaign, 1002 W. Green Street, Urbana, IL 61801, USA}
\author[0000-0002-3130-0204]{J.~Carretero}
\affiliation{Institut de F\'{\i}sica d'Altes Energies (IFAE), The Barcelona Institute of Science and Technology, Campus UAB, 08193 Bellaterra (Barcelona) Spain}
\author{M.~Costanzi}
\affiliation{INAF-Osservatorio Astronomico di Trieste, via G. B. Tiepolo 11, I-34143 Trieste, Italy}
\affiliation{Institute for Fundamental Physics of the Universe, Via Beirut 2, 34014 Trieste, Italy}
\author{L.~N.~da Costa}
\affiliation{Laborat\'orio Interinstitucional de e-Astronomia - LIneA, Rua Gal. Jos\'e Cristino 77, Rio de Janeiro, RJ - 20921-400, Brazil}
\affiliation{Observat\'orio Nacional, Rua Gal. Jos\'e Cristino 77, Rio de Janeiro, RJ - 20921-400, Brazil}
\author[0000-0001-8318-6813]{J.~De~Vicente}
\affiliation{Centro de Investigaciones Energ\'eticas, Medioambientales y Tecnol\'ogicas (CIEMAT), Madrid, Spain}
\author[0000-0002-0466-3288]{S.~Desai}
\affiliation{Department of Physics, IIT Hyderabad, Kandi, Telangana 502285, India}
\author[0000-0002-8357-7467]{H.~T.~Diehl}
\affiliation{Fermi National Accelerator Laboratory, P. O. Box 500, Batavia, IL 60510, USA}
\author{P.~Doel}
\affiliation{Department of Physics \& Astronomy, University College London, Gower Street, London, WC1E 6BT, UK}
\author[0000-0002-1894-3301]{T.~F.~Eifler}
\affiliation{Department of Astronomy/Steward Observatory, University of Arizona, 933 North Cherry Avenue, Tucson, AZ 85721-0065, USA}
\affiliation{Jet Propulsion Laboratory, California Institute of Technology, 4800 Oak Grove Dr., Pasadena, CA 91109, USA}
\author{S.~Everett}
\affiliation{Santa Cruz Institute for Particle Physics, Santa Cruz, CA 95064, USA}
\author[0000-0002-2367-5049]{B.~Flaugher}
\affiliation{Fermi National Accelerator Laboratory, P. O. Box 500, Batavia, IL 60510, USA}
\author[0000-0003-4079-3263]{J.~Frieman}
\affiliation{Fermi National Accelerator Laboratory, P. O. Box 500, Batavia, IL 60510, USA}
\affiliation{Kavli Institute for Cosmological Physics, University of Chicago, Chicago, IL 60637, USA}
\author[0000-0002-9370-8360]{J.~Garc\'ia-Bellido}
\affiliation{Instituto de Fisica Teorica UAM/CSIC, Universidad Autonoma de Madrid, 28049 Madrid, Spain}
\author[0000-0001-9632-0815]{E.~Gaztanaga}
\affiliation{Institute of Space Sciences (ICE, CSIC),  Campus UAB, Carrer de Can Magrans, s/n,  08193 Barcelona, Spain}
\affiliation{Institut d'Estudis Espacials de Catalunya (IEEC), 08034 Barcelona, Spain}
\author[0000-0003-3270-7644]{D.~Gruen}
\affiliation{SLAC National Accelerator Laboratory, Menlo Park, CA 94025, USA}
\affiliation{Kavli Institute for Particle Astrophysics \& Cosmology, P. O. Box 2450, Stanford University, Stanford, CA 94305, USA}
\affiliation{Department of Physics, Stanford University, 382 Via Pueblo Mall, Stanford, CA 94305, USA}
\author[0000-0002-4588-6517]{R.~A.~Gruendl}
\affiliation{National Center for Supercomputing Applications, 1205 West Clark St., Urbana, IL 61801, USA}
\affiliation{Department of Astronomy, University of Illinois at Urbana-Champaign, 1002 W. Green Street, Urbana, IL 61801, USA}
\author[0000-0003-3023-8362]{J.~Gschwend}
\affiliation{Laborat\'orio Interinstitucional de e-Astronomia - LIneA, Rua Gal. Jos\'e Cristino 77, Rio de Janeiro, RJ - 20921-400, Brazil}
\affiliation{Observat\'orio Nacional, Rua Gal. Jos\'e Cristino 77, Rio de Janeiro, RJ - 20921-400, Brazil}
\author[0000-0003-0825-0517]{G.~Gutierrez}
\affiliation{Fermi National Accelerator Laboratory, P. O. Box 500, Batavia, IL 60510, USA}
\author[0000-0002-6550-2023]{K.~Honscheid}
\affiliation{Center for Cosmology and Astro-Particle Physics, The Ohio State University, Columbus, OH 43210, USA}
\affiliation{Department of Physics, The Ohio State University, Columbus, OH 43210, USA}
\author[0000-0001-5160-4486]{D.~J.~James}
\affiliation{Center for Astrophysics $\vert$ Harvard \& Smithsonian, 60 Garden Street, Cambridge, MA 02138, USA}
\author[0000-0001-8356-2014]{E.~Krause}
\affiliation{Department of Astronomy/Steward Observatory, University of Arizona, 933 North Cherry Avenue, Tucson, AZ 85721-0065, USA}
\author[0000-0003-0120-0808]{K.~Kuehn}
\affiliation{Lowell Observatory, 1400 Mars Hill Rd, Flagstaff, AZ 86001, USA}
\affiliation{Australian Astronomical Optics, Macquarie University, North Ryde, NSW 2113, Australia}
\author[0000-0003-2511-0946]{N.~Kuropatkin}
\affiliation{Fermi National Accelerator Laboratory, P. O. Box 500, Batavia, IL 60510, USA}
\author[0000-0002-1134-9035]{O.~Lahav}
\affiliation{Department of Physics \& Astronomy, University College London, Gower Street, London, WC1E 6BT, UK}
\author[0000-0001-9856-9307]{M.~A.~G.~Maia}
\affiliation{Laborat\'orio Interinstitucional de e-Astronomia - LIneA, Rua Gal. Jos\'e Cristino 77, Rio de Janeiro, RJ - 20921-400, Brazil}
\affiliation{Observat\'orio Nacional, Rua Gal. Jos\'e Cristino 77, Rio de Janeiro, RJ - 20921-400, Brazil}
\author[0000-0003-0710-9474]{J.~L.~Marshall}
\affiliation{George P. and Cynthia Woods Mitchell Institute for Fundamental Physics and Astronomy, and Department of Physics and Astronomy, Texas A\&M University, College Station, TX 77843,  USA}
\author[0000-0002-8873-5065]{P.~Melchior}
\affiliation{Department of Astrophysical Sciences, Princeton University, Peyton Hall, Princeton, NJ 08544, USA}
\author[0000-0002-1372-2534]{F.~Menanteau}
\affiliation{Department of Astronomy, University of Illinois at Urbana-Champaign, 1002 W. Green Street, Urbana, IL 61801, USA}
\affiliation{National Center for Supercomputing Applications, 1205 West Clark St., Urbana, IL 61801, USA}
\author[0000-0002-6610-4836]{R.~Miquel}
\affiliation{Instituci\'o Catalana de Recerca i Estudis Avan\c{c}ats, E-08010 Barcelona, Spain}
\affiliation{Institut de F\'{\i}sica d'Altes Energies (IFAE), The Barcelona Institute of Science and Technology, Campus UAB, 08193 Bellaterra (Barcelona) Spain}
\author[0000-0002-6011-0530]{A.~Palmese}
\affiliation{Fermi National Accelerator Laboratory, P. O. Box 500, Batavia, IL 60510, USA}
\affiliation{Kavli Institute for Cosmological Physics, University of Chicago, Chicago, IL 60637, USA}
\author[0000-0002-2598-0514]{A.~A.~Plazas}
\affiliation{Department of Astrophysical Sciences, Princeton University, Peyton Hall, Princeton, NJ 08544, USA}
\author[0000-0002-9646-8198]{E.~Sanchez}
\affiliation{Centro de Investigaciones Energ\'eticas, Medioambientales y Tecnol\'ogicas (CIEMAT), Madrid, Spain}
\author{V.~Scarpine}
\affiliation{Fermi National Accelerator Laboratory, P. O. Box 500, Batavia, IL 60510, USA}
\author[0000-0001-9504-2059]{M.~Schubnell}
\affiliation{Department of Physics, University of Michigan, Ann Arbor, MI 48109, USA}
\author{S.~Serrano}
\affiliation{Institute of Space Sciences (ICE, CSIC),  Campus UAB, Carrer de Can Magrans, s/n,  08193 Barcelona, Spain}
\affiliation{Institut d'Estudis Espacials de Catalunya (IEEC), 08034 Barcelona, Spain}
\author[0000-0002-1831-1953]{I.~Sevilla-Noarbe}
\affiliation{Centro de Investigaciones Energ\'eticas, Medioambientales y Tecnol\'ogicas (CIEMAT), Madrid, Spain}
\author[0000-0002-3321-1432]{M.~Smith}
\affiliation{School of Physics and Astronomy, University of Southampton,  Southampton, SO17 1BJ, UK}
\author[0000-0002-7047-9358]{E.~Suchyta}
\affiliation{Computer Science and Mathematics Division, Oak Ridge National Laboratory, Oak Ridge, TN 37831}
\author[0000-0003-1704-0781]{G.~Tarle}
\affiliation{Department of Physics, University of Michigan, Ann Arbor, MI 48109, USA}

\collaboration{(DES Collaboration)}

\correspondingauthor{Alex Drlica-Wagner, Keith Bechtol}
\email{kadrlica@fnal.gov, kbechtol@wisc.edu}

\begin{abstract}
We report the results of a systematic search for ultra-faint Milky Way satellite galaxies using data from the Dark Energy Survey (DES) and Pan-STARRS1 (PS1).
Together, DES and PS1 provide multi-band photometry in optical/near-infrared wavelengths over $\roughly 80\%$ of the sky.
Our search for satellite galaxies targets $\roughly 25{,}000 \deg^2$ of the high-Galactic-latitude sky reaching a $10\sigma$ point-source depth of $\gtrsim 22.5$ mag in the $g$ and $r$ bands.
While satellite galaxy searches have been performed independently on DES and PS1 before, this is the first time that a self-consistent search is performed across both data sets.
We do not detect any new high-significance satellite galaxy candidates, while recovering the majority of satellites previously detected in surveys of comparable depth.
We characterize the sensitivity of our search using a large set of simulated satellites injected into the survey data.
We use these simulations to derive both analytic and machine-learning models that accurately predict the detectability of Milky Way satellites as a function of their distance, size, luminosity, and location on the sky.
To demonstrate the utility of this observational selection function, we calculate the luminosity function of Milky Way satellite galaxies, assuming that the known population of satellite galaxies is representative of the underlying distribution. 
We provide access to our observational selection function to facilitate comparisons with cosmological models of galaxy formation and evolution.
\end{abstract}

\reportnum{DES-2019-0468}
\reportnum{FERMILAB-PUB-19-604-AE}

\section{Introduction}

Faint dwarf galaxies dominate the universe by number, yet a precise census of these objects remains challenging, due to the limited sensitivity of observational surveys.
Dwarf galaxies with stellar mass $\lesssim 10^6 \Msun$ have only been identified within the Local Volume (distances of a few Mpc), either in isolation or as satellites of larger galaxies \citep[\eg,][]{Martin:2013, Muller:2015, Carlin:2016, Smercina:2018, Crnojevic:2019}. 
At even lower masses, the census of ultra-faint satellites is incomplete, even within the Milky Way halo.
Despite significant observational challenges, the demographics of ultra-faint dwarf galaxies offer a unique window into feedback processes in galaxy formation \citep[e.g.][]{Maschenko07114803,Wheeler150402466,Wheeler181202749,Munshi181012417,Agertz190402723}, reionization and the first stars \citep[\eg,][]{Bullock0002214,Shaprio2004,Weisz:2014a,Weisz:2014b,Boylan-Kolchin:2015,Ishiyama160200465,Weisz:2017,Tollerud2018,Graus:2019,Katz190511414}, and the nature of dark matter \citep[e.g.,][]{Bergstrom:1998,Spekkens:2013,Malyshev:2014,Ackermann:2015,Geringer-Sameth:2015,Bullock:2017,Clesse:2018,Nadler:2019b}.

\begin{figure*}
\includegraphics[width=\textwidth]{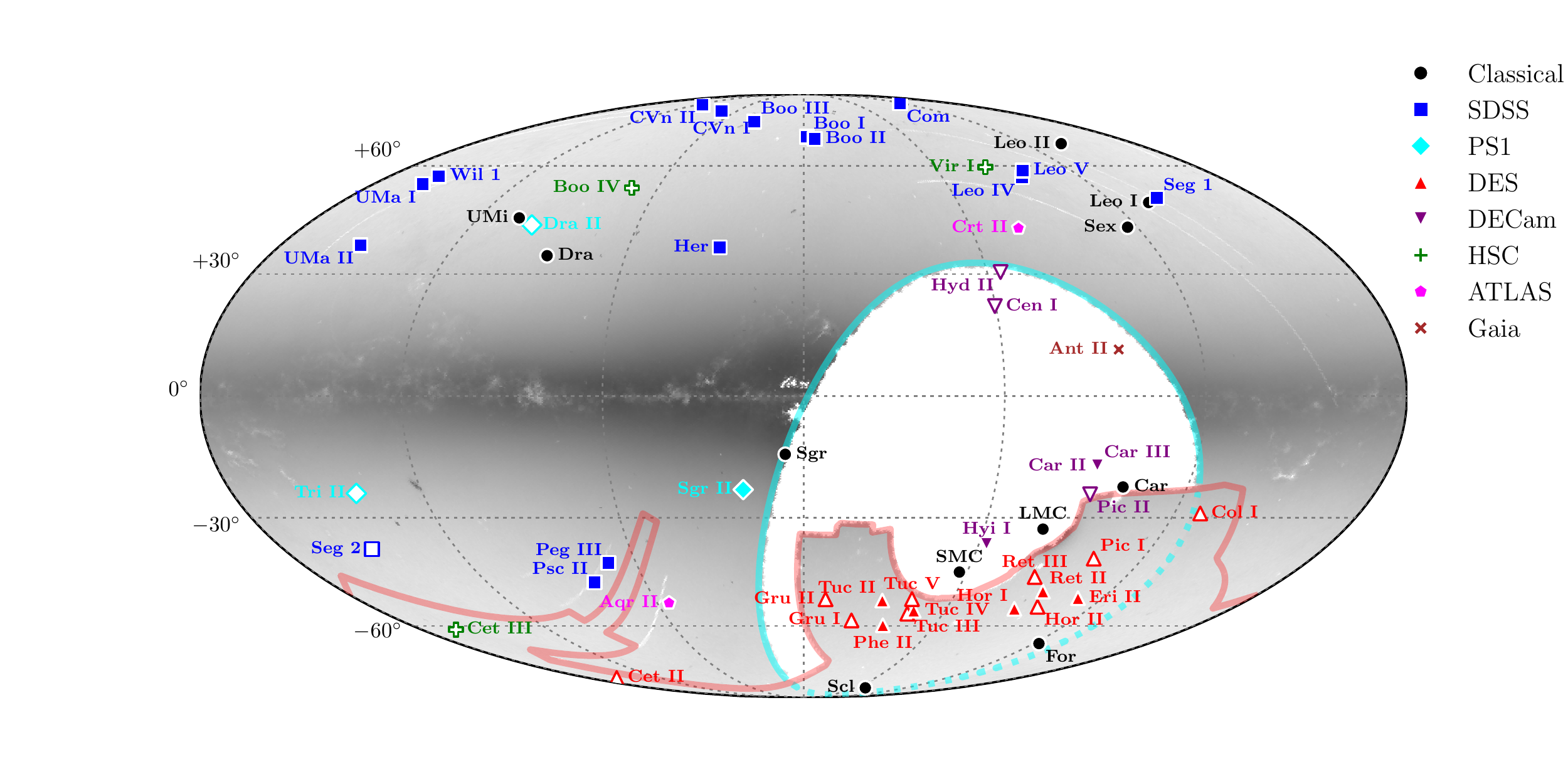}
\caption{Equal-area Mollweide projection of the density of stars (log scale) observed with $r < 22 \magn$ over the DES Y3A2 and PS1 DR1 footprints (red and cyan borders, respectively).
Filled markers indicate kinematically confirmed Milky Way satellite galaxies, and unfilled markers indicate satellite galaxy candidates that have not yet been kinematically confirmed.
We mark classical Milky Way satellites (black circles), and satellites discovered by SDSS (blue squares), PS1 (cyan diamonds), DES (red upright triangles), other DECam surveys (purple inverted triangles), HSC SSP (green pluses), VLT ATLAS (magenta pentagons), and \Gaia (brown crosses).
}
\label{fig:skymap_zoom}
\end{figure*}

The lowest-luminosity satellite galaxies are detected in optical imaging surveys as arcminute-scale statistical overdensities of individually resolved stars.
Beginning with the Sloan Digital Sky Survey (SDSS), digitized wide-area multi-band optical imaging surveys---combined with automated search algorithms---have greatly increased the known population of Milky Way satellites \citep{2005AJ....129.2692W,2005ApJ...626L..85W,2006ApJ...650L..41Z,2006ApJ...643L.103Z,2006ApJ...647L.111B,2007ApJ...654..897B,2008ApJ...686L..83B,2009MNRAS.397.1748B,2010ApJ...712L.103B,2006ApJ...645L..37G,Grillmair:2009,2006ApJ...653L..29S,2007ApJ...656L..13I,2007ApJ...662L..83W,Kim:2015d}.
More recently, searches using data from the Dark Energy Survey \citep[DES;][]{Bechtol:2015, Koposov:2015, Kim:2015c, Drlica-Wagner:2015, Luque:2016}, other DECam surveys \citep[\eg, SMASH, MagLiteS, and DELVE;][]{martin_2015_hydra_ii,Drlica-Wagner:2016hwk,Torrealba:2018a,Koposov:2018,Mau:2019b}, ATLAS \citep{Torrealba:2016a,Torrealba:2016b}, Pan-STARRS1 \citep[PS1;][]{Laevens:2015a,Laevens:2015b}, and \Gaia \citep{Torrealba:2019} have further increased the sample of confirmed and candidate satellites to more than 50 (\figref{skymap_zoom}).

Both observational and theoretical arguments suggest that the current census of Milky Way satellite galaxies is incomplete.
From an observational standpoint, this incompleteness is demonstrated by the continued discovery of fainter, more distant, and lower surface brightness systems.
For example, the first $\roughly 700 \deg^2$ of deep imaging with the Hyper Suprime-Cam Strategic Survey Program (HSC SSP) has revealed three new satellites at sufficiently low luminosities and large heliocentric distances that they escaped detection by earlier overlapping surveys \citep{Homma:2016,Homma:2017,Homma:2019}.
Moreover, several recently discovered Milky Way companions (\eg, Crater~II, Virgo~I, Aquarius~II, Cetus~III, Antlia~II, and Bo{\"o}tes~IV) are lower surface brightness than most ultra-faint dwarfs discovered in the SDSS era, implying that the current generation of surveys and search techniques are sensitive to systems that were previously undetectable.
Searches using compact spatial kernels and a wider variety of stellar population ages and metallicities have revealed diverse Milky Way substructures \citep{Torrealba:2018b}, and precise proper motion information for billions of nearby stars provided by \Gaia has enlarged the sample of extremely low-surface-brightness satellites \citep{Torrealba:2019}. 

Theoretical predictions for the smallest galaxies have advanced hand in hand with observations.
Since galaxy formation is a nonlinear process, numerical simulations have long been used to predict the population statistics of these objects.
Early simulations that resolved dark matter substructure within Milky Way-mass halos predicted far more surviving dark matter subhalos than the number of observed satellites \citep{Klypin9901240,Moore9907411}. 
This mismatch, dubbed the ``missing satellites problem,'' simply reflects the fact that mapping subhalos in dark-matter-only simulations to observed satellites is nontrivial.
In particular, reionization and stellar feedback drastically suppress dwarf galaxy formation in low-mass halos \citep[\eg,][]{Bullock0002214,Somerville0107507,Brown14100681}, and tidal interactions with the Galactic disk are expected to disrupt a significant number of systems \citep[\eg,][]{Garrison-Kimmel:2017,Kelley:2018,Nadler:2018}. 
Semi-empirical models that account for these effects---along with realistic satellite detection criteria---find that the observed satellite population is consistent with cold, collisionless dark matter \citep[\eg,][]{Kim:2017,Jethwa:2018,Newton:2018,Nadler:2019a,Nadler:2019b,Bose:2019}. 
Likewise, hydrodynamic simulations that self-consistently model galaxy formation in a cosmological context produce luminosity functions and radial distributions of satellites that are broadly consistent with observations of the Milky Way system \citep[\eg,][]{Wetzel160205957,GarrisonKimmel180604143,Samuel:2020}.
In concert, extremely high-resolution simulations of isolated ultra-faint systems suggest that low-mass dwarfs may be abundant \citep{Wheeler181202749}.

Historically, the primary means of comparing Milky Way satellite observations to simulations has been through the total satellite luminosity function (\ie, the total number of satellites within the virial radius of the Milky Way halo as a function of satellite luminosity). 
Typically, an observational selection function is built to predict the detectability of a satellite as a function of heliocentric distance, size, and luminosity.
This type of analysis was pioneered by \citet{Koposov:2008} and \citet{Walsh:2009}, who used simulations to characterize the satellite detection efficiency in SDSS, analyzing $\roughly 8{,}000 \deg^2$ from SDSS DR5 and $\roughly9{,}500 \deg^2$ from SDSS DR6, respectively.
The total luminosity function was derived by correcting the observed satellite population for observational selection effects, and the result was compared to cosmological predictions. 
Recently, several studies have begun to utilize more advanced model inference techniques that require a simple yet comprehensive mechanism to predict the detectability of a satellite \citep{Jethwa:2018,Newton:2018,Nadler:2019a}.
However, these studies have been limited by the lack of rigorous estimates for the selection functions of modern surveys.

In this paper, we present a systematic search for Milky Way satellites and a detailed quantitative measurement of the observational selection function for modern surveys. 
In particular, we performed an updated search for Milky Way satellites by applying two independent search algorithms to $\roughly 5{,}000 \deg^2$ of data from DES \cite{DR1:2018} and $\roughly 30{,}000 \deg^2$ of data from PS1 \citep{Chambers:2016}. 
After quality cuts, our analysis covers approximately three times the sky area analyzed by \citet{Koposov:2008} and \citet{Walsh:2009}.
The DES data were collected during the first three years of survey operations and cover much of the southern Galactic cap \citep[DES Y3A2;][]{DR1:2018, Shipp:2018}. 
When compared to previous DES satellite searches \citep[\ie,][]{Drlica-Wagner:2015}, DES Y3A2 has $\roughly 50\%$ more exposure time, more homogeneous coverage, more accurate photometric calibration, and more efficient star--galaxy classification \citep[\eg,][]{Burke:2018,DR1:2018}.
To extend the coverage of our analysis to the northern hemisphere, we also apply our search algorithms to publicly available data from the first data release of PS1 \citep[PS1 DR1;][]{Chambers:2016}.
Note that in most regions of the sky at high Galactic latitude, the number density of background galaxies exceeds that of foreground Milky Way stars at magnitudes $r \gtrsim 22$. 
Accordingly, this analysis represents a systematic search over $\roughly 75\%$ of the high-Galactic-latitude sky reaching depths at which the stellar sample is limited primarily by star--galaxy confusion, rather than object detection \citep[\eg,][]{Fadely:2012}. 

We quantify the observational selection function of our search to facilitate direct comparisons between the observed luminosity function and predictions from simulations. 
We simulate the resolved stellar populations of $10^5$ ($10^6$) satellites and inject simulated stars into the DES Y3A2 (PS1 DR1) data at the catalog level.
These simulations span a range of absolute magnitudes, heliocentric distances, physical sizes, ellipticities, position angles, ages, and metallicities.
We run our search algorithms on each simulated satellite and find that the detectability of a satellite can be well described by its absolute magnitude, heliocentric distance, physical size, and local stellar density.
We derive both analytic and machine-learning models that predict the detectability of a satellite as a function of these parameters.

The observational selection function derived in this paper can be used to test models that predict the abundance and properties of Milky Way satellites.
As an illustrative example, we use our observational selection function to derive the total luminosity function of Milky Way satellites solely based on the properties of the observed population.
In a companion paper \citep[][hereafter Paper II]{PaperII}, we use high-resolution numerical simulations (including a model for the effects of baryons) to build a more rigorous model of the observed satellite population and to constrain models of galaxy formation.
In deriving the observational selection function, we have intentionally set a high threshold for detection in order to provide a clean interpretation of the resulting satellite populations. 
The investigation of lower significance candidates is left to future work.

This paper is structured as follows.
In \secref{pipeline}, we provide a high-level overview of our simulation and analysis pipeline.
The subsequent sections provide more detail on the survey data sets (\secref{data}), our catalog-level simulations (\secref{simulations}), and the satellite search algorithms (\secref{algorithms}).
In \secref{results}, we present the results of our search on the DES and PS1 data. 
The resulting observational selection functions derived from simulations are presented in \secref{ssf},  and our simple luminosity function inference is presented in \secref{lf}.  
We conclude in \secref{conclusions}.
Our models for the observational selection functions of DES and PS1 are publicly available online.\footnote{\url{https://github.com/des-science/mw-sats}}

\section{Analysis Overview}
\label{sec:pipeline}

In this section, we summarize the key components of the simulation and data analysis pipeline used to derive the observational selection function for Milky Way satellites.
We applied two distinct algorithms to search for satellite galaxies in photometric catalog data from DES and PS1.
To evaluate the sensitivity of our search, we embedded simulated satellite galaxies into these data and attempted to recover them with the same search algorithms.
By self-consistently analyzing the data and simulations, we accurately characterized both the population of observed satellites and the population of satellites that remain undetected due to the limited sensitivity of our observations.

To generate realistic satellite galaxy simulations, we empirically modeled the survey coverage and photometric response of DES Y3A2 and PS1 DR1.
We characterized the coverage, depth, completeness, and photometric measurement uncertainties as a function of sky location for each survey.
We then simulated stellar catalogs for satellites covering a large range of physical properties, including sky location, luminosity, heliocentric distance, physical size, ellipticity, age, and metallicity.
For each satellite, we generated a Poisson realization of the observable stellar distribution, simulating the position, flux, and photometric uncertainty of each star.
These simulated satellites were injected one-by-one into the survey data sets, and two search algorithms were run at the location of each injected system.
The analysis of these simulations produces a multi-dimensional vector containing the detection significance of each simulated satellite as a function of its intrinsic properties (\eg, luminosity, distance, and physical size) and global survey properties (\eg, survey depth, coverage, and local foreground stellar density).
We refer to the mapping between satellite properties and satellite detectability as the observational selection function.

In parallel, we performed an untargeted search of DES Y3A2 and PS1 DR1 without any embedded simulations. 
This search produced a set of stellar overdensity ``seeds'' of varying significance.
The most significant seeds are associated with physical systems reported in the literature, while less significant seeds can be attributed to statistical fluctuations, artificial variations in the stellar density due to survey systematics, and sub-threshold physical systems.
We characterized the distribution of detection significances for the collection of seeds and defined a conservative detection threshold that recovers a large fraction of systems that were discovered in surveys of comparable depth.
This {\it a posteriori} definition of a detection threshold was required to deal with systematic artifacts that contaminated the population of seeds and made it impossible to choose a statistical threshold {\it a priori}.
However, by self-consistently applying the same detection threshold to the population of simulated satellites, we can determine the selection efficiency for any detection threshold and satellite properties.
In this paper, we are primarily concerned with the demographics of the satellite galaxy population rather than detecting new, low-significance candidates, and therefore we set our significance threshold to yield a pure sample of ``detected'' satellites.

The detected population of satellites and the observational selection function can be combined to derive the Milky Way satellite galaxy luminosity function.
Analyzing the results of the simulations directly can be cumbersome and computationally intensive, while representing the observational selection function with a simple analytic relationships discards some information.
Therefore, to simplify the application of the selection function while retaining detailed information, we trained a gradient-boosted decision tree classifier that takes as input characteristics of a satellite (\eg, size, luminosity, distance, and local stellar density) and outputs a probability that the satellite would be detected.
When applying this classifier, we combine the satellite properties with the global geometric characteristics of each survey in the form of \healpix maps of survey coverage.

\begin{figure*}
\hspace{20mm}
  \includegraphics[width=0.9\textwidth]{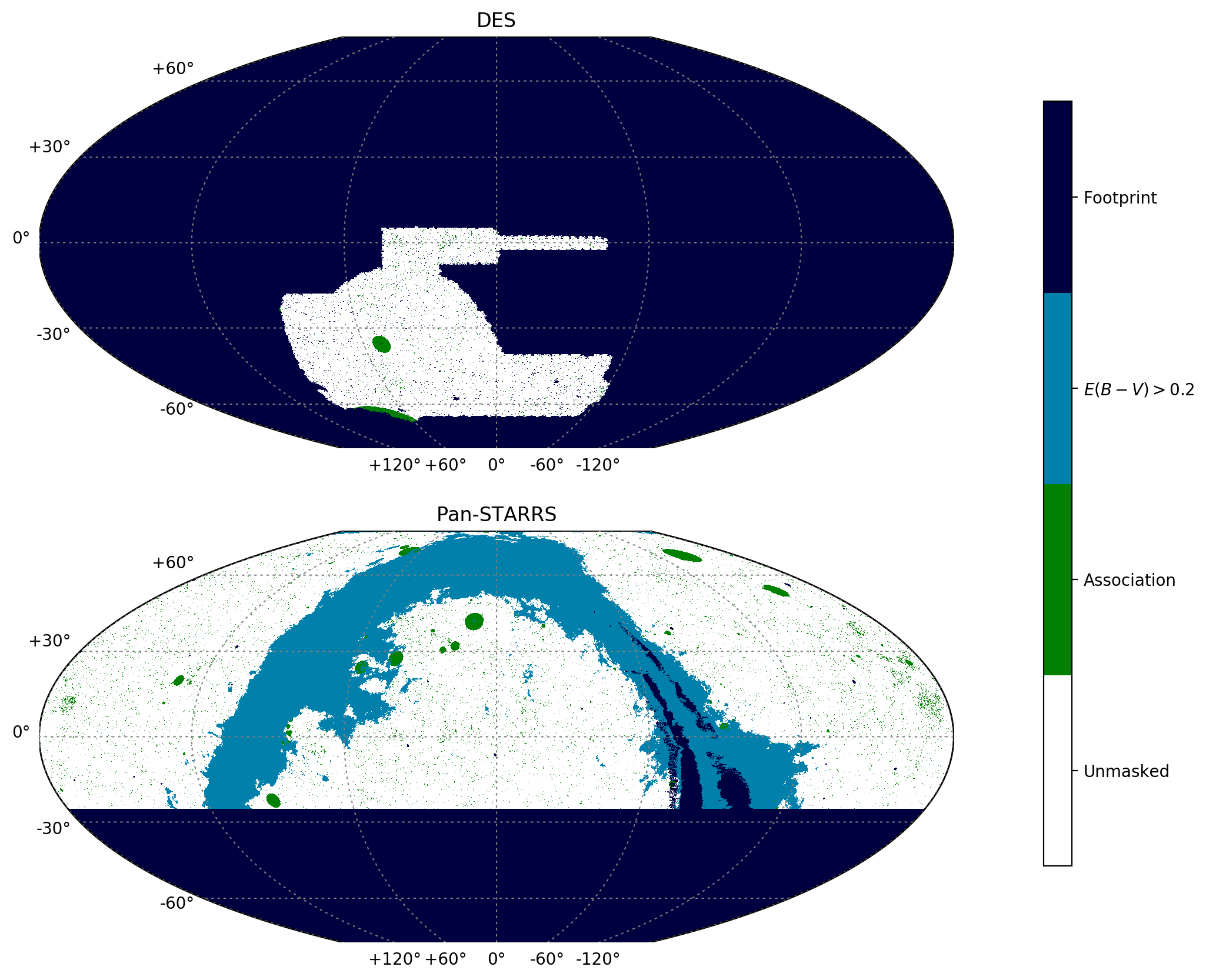}
\caption{McBryde-Thomas flat polar quartic projection of the geometric masks applied to the DES and PS1 satellite searches.
Colored regions are masked from our analysis either because they lie outside the DES or PS1 survey footprint (dark blue), have interstellar extinction $E(B-V) > 0.2 \magn$ in the \cite{Schlegel:1998} dust map (blue), or are associated with previously known stellar overdensities, galaxy clusters, or bright stars (green).
The total unmasked area is \CHECK{$4{,}844 \deg^2$ for DES, $21{,}123 \deg^2$ for PS1, and $24{,}343 \deg^2$} for the combined search.
}
\label{fig:masks}
\end{figure*}

\section{Data Set}\label{sec:data}

Data from DES Y3A2 and PS1 DR1 cover $\roughly 5{,}000\deg^2$ and $\roughly 30{,}000\deg^2$ of the celestial sphere, respectively (\figref{masks}).
The deep, multi-band, optical/near-infrared imaging of these surveys provides the photometric, astrometric, and morphological measurements necessary to separate stellar overdensities in the Milky Way halo from Milky Way field stars and unresolved background galaxies.
In this section, we describe the selection of high-quality stellar samples for each of these surveys, the characterization of the survey geometry, and determination of survey response as a function of location on the sky.   
Additional technical details on our selections are provided in \appref{ps1}.

\subsection{DES Y3A2}
\label{sec:des}

DES is a broadband optical/near-infrared imaging survey of the southern Galactic cap using the Dark Energy Camera \citep[DECam;][]{Flaugher:2015} mounted at the prime focus of the 4-m Blanco telescope at the Cerro Tololo Inter-American Observatory (CTIO).
Here, we analyze data from the first three years of DES operations \citep{Diehl:2016}.
The DES Y3A2 imaging data serves as the basis for the first DES public data release \citep[DES DR1;][]{DR1:2018} and consists of $\roughly 45{,}000$ wide-area survey exposures.
Details of the DES image reduction and catalog generation can be found in \citet{Morganson:2018}, while more details on the internal DES Y3A2 data set can be found in \citet{Sevilla:2019}.

\paragraph{Photometry}
The internal DES Y3A2 object catalogs augment DES DR1 with additional multi-band, multi-epoch, forced photometry, which provides significantly improved photometric and morphological measurements of faint objects.
These catalogs were generated in two steps.
First, individual sources were detected in $riz$ coadded images using \sextractor \citep{1996A&AS..117..393B}, with a detection threshold of $S/N \sim 10$ \citep{Morganson:2018}.
This coadd object catalog was then used as input to the \ngmix multi-band, multi-epoch fitting routine, which performs a simultaneous fit of source parameters across the set of individual {\it griz} images for each object \citep{Sheldon:2014,Drlica-Wagner:2018}. 
\ngmix is run in two configurations: (1) fits are performed on single objects while masking nearby neighbors, referred to as the ``single object fit'' (SOF), and (2) fits are performed iteratively on groups of objects, referred to as the ``multi-object fit'' (MOF).
The treatment of the PSF on an image-by-image basis substantially improves point-source photometry and star--galaxy separation \citep{Sevilla:2018}.
The relative top-of-the-atmosphere photometric accuracy across the DES footprint is estimated to be $< 7\mmag$ from a comparison to \Gaia \citep{Burke:2018}.
Furthermore, DES Y3A2 includes SED-dependent chromatic corrections for each object based on an initial evaluation of the stellar spectral type \citep{Li:2016,Sevilla:2019}.

\paragraph{Coverage} 
The observational coverage of DES Y3A2 was assembled in a vectorized format using \mangle \citep{Hamilton:2004,Swanson:2008}.
The \mangle representation accounts for missing coverage at the boundary of the survey footprint, as well as gaps associated with saturated stars, bleed trails, and other instrument signatures. 
These vectorized maps were then converted into a subsampled $\nside = 4096$ \healpix map with a resolution of $\roughly 0.74 \amin^2$ per pixel \citep{Drlica-Wagner:2018}.
We restrict the survey footprint to \healpix pixels where the $griz$ sky coverage fraction is greater than 0.5, resulting in a total effective solid angle before masking of \CHECK{$4{,}945 \deg^2$}.

\paragraph{Depth} 
The depth of DES Y3A2 was estimated for each band in each $\nside=4096$ \healpix pixel.
This involved combining the \code{mangle} maps with additional survey characteristics following the procedure developed in \citet{Rykoff:2015}, as described in Section 7.1 of \citet{Drlica-Wagner:2018}.
In brief, we trained a random forest classifier that combined survey characteristics, such as coverage, seeing, and sky brightness, to estimate the $10\sigma$ limiting magnitude in each pixel.
The $10\sigma$ SOF \code{CM\_MAG} depth for DES Y3A2 data in regions with $E(B-V) < 0.2$ is $g = 23.9$ and $r = 23.7$. 
When simulating and analyzing satellites in DES Y3A2, we incorporated the (small) variations in depth over the footprint.

\paragraph{Reddening} 
We followed the procedure described in Section 4.2 of \citet{DR1:2018} to correct for interstellar extinction. 
We applied an additive correction to each measured magnitude of $A_b = R_b \times E(B-V)$, where $E(B-V)$ comes from \citet[][SFD]{Schlegel:1998}.
We use the $R_b$ coefficients from \citet{DR1:2018}---specifically $R_g = 3.186$ and $R_r = 2.140$. 
These fiducial coefficients are derived using the \citet{Fitzpatrick:1999} reddening law with $R_V = 3.1$ and incorporate the renormalization of the SFD reddening map ($N = 0.78$) suggested by \citet{Schlafly:2010}.
Hereafter, all DES Y3A2 magnitudes are extinction corrected.

\paragraph{Star--Galaxy Separation}
The star--galaxy separation efficiency of previous DES data sets was estimated to be $>95\%$ complete for stars at $i < 22$ \citep{Drlica-Wagner:2018}. 
\citet{Sevilla:2018} showed that the efficiency of star--galaxy separation could be considerably improved by using multi-epoch morphological measurements. 
We assessed the efficiency of star--galaxy classification in DES Y3A2 by comparing to overlapping deep data from the HSC SSP \citep{HSC_DR1}. 
Accounting for both object detection and classification, we found that the DES Y3A2 SOF classifier\footnote{We use $0 \leq \var{EXTENDED\_CLASS\_MASH\_SOF} \leq 2$ to define the stellar sample, balancing stellar completeness and galaxy contamination. Details on the performance of star--galaxy classification for DES Y3A2 will be presented in \citet{Sevilla:2019}.} achieves $>90\%$ completeness for stars down to a magnitude limit of $r \sim 23.5 \magn$.
The stellar completeness of ground-based surveys such as DES and PS1 is largely dominated by the efficiency of star--galaxy classification rather than the point-source detection limit. 

\subsection{PS1 DR1}
\label{sec:ps1}

Our PS1 data set consists of data from the first public data release of the PS1 $3\pi$ Survey \citep{Chambers:2016}. 
PS1 DR1 is assembled from images taken by the PS1 Gigapixel Camera \#1 \citep{Tonry:2008} mounted on the 1.8-m PS1 telescope at Haleakala Observatories on the island of Maui, Hawai`i.
PS1 DR1 covers the northern and equatorial sky with $\dec > -30 \deg$ in five optical bands, $\gp,\rp,\ip,\zp,\yp$ \citep{Tonry:2012}. 
The DES and PS1 $g$, $r$, and $i$ bands are similar enough that we drop the ``P1'' subscript, though we analyzed each survey in its respective filter system.

\paragraph{Photometry}
We selected PS1 DR1 objects from \code{ForcedMeanObject{\allowbreak}View} with \code{qualityFlag \& 16 > 0}, \code{nDetections > 0}  and \code{nStackDetections > 1}.
We removed duplicate objects from the catalogs by selecting only objects that are the primary detection, \code{detectPrimary = 1}.
We made several additional quality cuts based on the \code{InfoFlag} and \code{InfoFlag2} variables to remove objects where the photometric fit failed, objects that were likely to be defects, objects with too few points measured to derive an elliptical contour, and objects where all model fits failed.
The full set of selection criteria can be found in \appref{ps1}.
These criteria were validated by comparing against catalogs derived from the HSC SXDS ultra-deep field. 
We found that our cuts retain the majority of PS1-HSC matched objects while significantly reducing the incidence of spurious objects in the PS1 data.
We converted the measured fluxes from PS1 DR1 into magnitudes by applying a stack zero-point of 8.9 (\ie, $\magn = -2.5 \log_{10}({\rm flux}) + 8.9$).
When PS1 DR1 reports negative flux values, we set the corresponding magnitude and magnitude uncertainty to a sentinel value of -999.
Since we are using $(g-r)$ colors for our search, we selected only objects with measured PSF magnitudes in both the $g$ and $r$ bands.

\paragraph{Coverage} 
We approximated the coverage of PS1 empirically using the full PS1 DR1 catalog prior to any star--galaxy separation or photometric cuts. 
We define the PS1 footprint as the set of $\nside = 2048$ ($\roughly 2.95 \amin^2$) \healpix pixels that contain any PS1 object.\footnote{The choice of $\nside = 2048$ was determined empirically, as the highest resolution that did not contain a significant number of pixels that lacked objects, but were clearly covered by PS1 (as determined by visual inspection of the coadd images).}
This coverage map is not strictly accurate, since some \healpix pixels that contain objects are not fully covered by the survey while other \healpix pixels are covered by the survey yet contain no objects.
However, at the level of accuracy necessary for our search algorithms, this coverage map is sufficient to avoid significantly biasing estimates of the local stellar density.
We converted from $\nside = 2048$ to $\nside = 4096$ (\ie, setting the value of each nested subpixel to the value of its parent) for use by the analysis algorithms.
The total area of the PS1 DR1 footprint that we consider before masking is $29{,}343 \deg^2$.
 
\paragraph{Depth} 
We estimated the photometric depth of PS1 DR1 by interpolating the median magnitude uncertainty as a function of magnitude for a set of low-reddening, high-Galactic-latitude regions.
We determined the magnitude at which the median magnitude uncertainty is 0.1085, corresponding to the $10\sigma$ detection limit.
We found typical $10\sigma$ magnitude limits for PS1 DR1 of $g = 22.5$ and $r = 22.4$. 
These depth estimates agree with those of \citet{Chambers:2016}, who estimate that the PS1 DR1 catalog retains 98\% completeness at $g,r,i \sim 22.5$ with a spatial variation of $\pm 0.25 \magn$ \citep[see figure 17 of][]{Chambers:2016}. 
We assumed a constant magnitude limit for PS1 DR1; however, variable interstellar extinction introduces a spatially dependent intrinsic magnitude limit for stars (\ie, we are less sensitive in regions of high extinction).
We have found that stars fainter than the $10\sigma$ magnitude limit can contribute significantly to the detectability of faint satellites.
For this reason, we used a magnitude limit of $g,r = 23$ when performing the likelihood-based search (\secref{ugali}).
In the spatial matched-filter analysis (\secref{simple}), which uses a binary selection in color--magnitude space rather than a likelihood-based weighting of photometric uncertainties, we applied a signal-to-noise threshold that limits the completeness of faint stars but better controls the number of spurious seeds returned by the algorithm.

\paragraph{Reddening} 
We corrected the PS1 DR1 measured magnitudes for interstellar extinction following the procedure described in \citet{Schlafly:2011}.
$E(B-V)$ values were calculated from each object by performing a bilinear interpolation to the maps of \citet{Schlegel:1998}.
We then applied $R_b$ coefficients from table 6 of \citet{Schlafly:2011}, assuming the reddening law of \citet{Fitzpatrick:1999} with $R_V = 3.1$. 
For PS1 DR1, this corresponds to $R_g = 3.172$ and $R_r = 2.271$ for the $g$ and $r$ bands, respectively.
We note that these $R_b$ values include a renormalization factor, as suggested by \citet{Schlafly:2010}. 
Hereafter, all PS1 DR1 magnitudes are extinction corrected unless explicitly stated otherwise.

\paragraph{Star--Galaxy Separation} 
We performed star--galaxy separation by comparing the measured $i$-band PSF and adaptive aperture magnitudes, \code{iFPSFMag} and \code{iFKronMag}, respectively.
The choice of $i$-band was motivated by the superior PSF and depth in this band.
Our primary cut required that the measured PSF and aperture magnitudes agree, $\code{iFPSFMag} - \code{iFKronMag} < 0.05$. 
However, we found that the PS1 PSF fit often fails in dense stellar regions. 
To retain sensitivity in these regions, we also included objects where $\code{iFPSFMag} = -999$ or $\code{iFPSFMag} - \code{iFKronMag} > 4$.
By comparing to HSC SSP \citep{HSC_DR1}, we find that our PS1 DR1 stellar sample is $>90\%$ complete down to a magnitude of $r \sim 21.7$. 

\begin{figure*}[th]
\includegraphics[width=1.\textwidth]{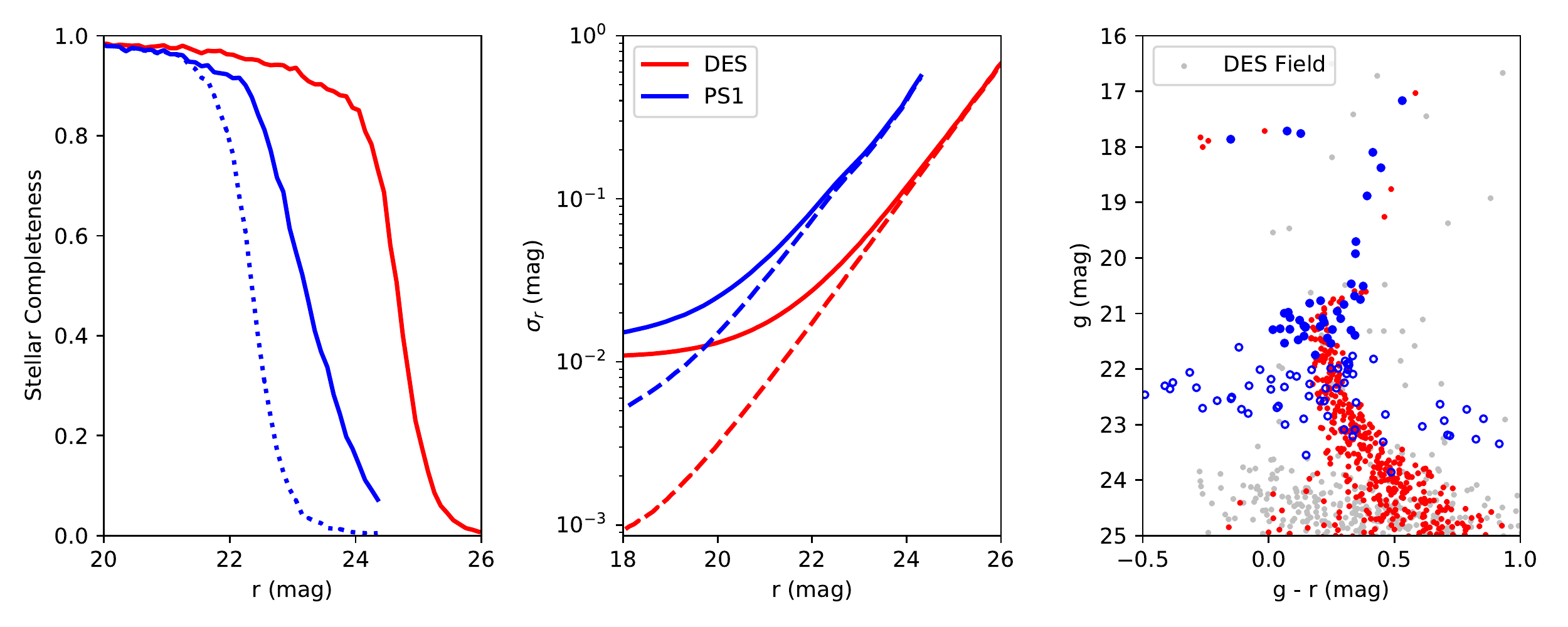}
\caption{\label{fig:photo_err}
(Left) Stellar completeness models for the DES Y3A2 and PS1 DR1 data sets. 
The effects of both object detection and star--galaxy classification are included. 
The dotted and solid curves for PS1 indicate stellar completeness with and without the S/N selection criteria, respectively.
(Center) Photometric uncertainty models for both data sets. Dashed lines correspond to statistical uncertainty alone, while the solid lines impose a minimum photometric uncertainty of $0.01 \magn$.
(Right) Color--magnitude distribution of stars in a simulated satellite with heliocentric distance $D = 28.8 \kpc$ and stellar mass $M_\star = 880 \Msun$ as realized in the DES (red) and PS1 (blue) surveys. 
Open circles for PS1 indicate stars excluded by the S/N selection criteria. 
DES field stars in a region of 0.1 deg radius are shown for comparison (gray). 
}
\end{figure*}

\section{Satellite Simulations}
\label{sec:simulations}

We simulated Milky Way satellite galaxies with a wide range of properties to accurately quantify the detection efficiency of our algorithms. 
We randomly sampled values of stellar luminosity, heliocentric distance, physical size, ellipticity, and position angle from the ranges described in \tabref{sim_params}.
Stellar photometry was simulated based on stellar isochrones from \citet{Bressan:2012}, selected from a range of ages and metallicities characteristic of observed satellites (\tabref{sim_params}).
Photometric error models were derived for each survey and were used to assign a photometric uncertainty to each star and randomize the measured photometry relative to the deterministic photometry provided from the isochrone.
The simulated satellites were randomly assigned spatial locations in a region that slightly overcovered each of the survey footprints.
The population of simulated satellites was not intended to mimic any realistic satellite population; rather, it was intended to cover the range of parameter space where variations in detection efficiency occur.

Satellites were simulated at the catalog level as collections of individually resolved stars. 
To generate realistic catalogs, we began with a probabilistic model for the spatial and flux distributions of stars in each satellite.
We sampled the spatial distribution of stars according to a Plummer profile \citep{Plummer:1911}, which has been found to be a good description of known Milky Way satellite galaxies \citep[\eg,][]{Simon:2019}.\footnote{Assuming a different spatial profile (\eg, an exponential profile) has little impact on detectability; however, see \citet{Moskowitz:2019} for a detailed analysis of alternative spatial kernels.}
We use $a_h$ to indicate the elliptical semi-major axis containing half the light (arcmin) and $r_h = a_h \sqrt{1-e}$ to represent the azimuthally averaged half-light radius (arcmin), where $e$ is the ellipticity.
$a_{1/2}$ and $r_{1/2}$ represent the equivalent quantities as projected physical lengths (pc) at the heliocentric distance, $D$.

The initial masses of satellite member stars were drawn from a \citet{Chabrier:2001} initial mass function (IMF), which has been found to be a reasonable description of known satellite galaxies \citep{Simon:2019}. 
Initial stellar masses were used to assign current absolute magnitudes from a \citet{Bressan:2012} isochrone.
When sampling from the IMF, the lower mass bound was set to the hydrogen-burning limit of 0.08\Msun and the upper bound was set by the star with the largest initial mass in the evolved isochrone (white dwarfs are ignored).
Using the \citet{Bressan:2012} isochrones, we transformed from initial stellar mass to current absolute magnitude in the $g$ and $r$ bands for each survey, and then to apparent magnitudes using the distance modulus of the simulated satellite.
We applied interstellar extinction to the apparent magnitudes of each simulated star using the same reddening coefficients described in \secref{data}.

We estimated the photometric uncertainty on the simulated stellar magnitudes based on the depth of the survey at the location of each star according to the formula
\begin{equation}
\label{eqn:photo_error}
\sigma_m = 0.01 + 10^{f(\Delta m)}.
\end{equation}
The function, $f(\Delta m)$, maps the difference between the apparent magnitude of a star and the $10\sigma$ survey magnitude limit at the location of the star, $\Delta m = m_{\rm lim} - m$, to the median magnitude uncertainty.
We derive $f(\Delta m)$ for each survey by calculating the median magnitude uncertainty as a function of magnitude and magnitude limit.
In the middle panel of \figref{photo_err}, we plot our photometric uncertainty model as a function of $r$-band magnitude, given the characteristic depth of DES ($r_{\rm lim} = 23.7$) and PS1 ($r_{\rm lim} = 22.4$).

To assess the sensitivity of our search algorithms, we inserted simulated stellar catalogs for each satellite into the real data and ran our satellite search algorithms at the location of each injected satellite.
We simulated stellar catalogs for $10^5$ ($10^6$) satellites in the DES (PS1) footprint.\footnote{On average, the DES simulations generate many more member stars per satellite due to the deeper DES imaging (\figref{photo_err}). This makes it computationally challenging to simulate more DES satellites.}
To make economical use of compute time and simulated data volume, satellites with high surface brightness $\mu < 23.5 \magn \asec^{-2}$ and $> 10^3$ detected stars brighter than $g = 22$ were not fully simulated and were instead assumed to be detected if they reside within the geometric survey coverage masks (see \appref{mu_ng22} for details).
We record the detection significance of each simulated satellite, along with metadata about the survey characteristics at the injected location.

When analyzing the simulated satellites, we use the same configuration that was used to search the real data.
However, to save on computational time, we fixed the spatial location and distance modulus of our analysis to the value of the search grid that best matched the location and distance of the simulated satellite.
This yields a conservative estimate of the detection significance, since we are ignoring the possibility that background fluctuations could slightly enhance the detection significance at other locations or distances.
To assess the impact of this choice, we freed the distance modulus for a small set of simulated satellites and found that the detection probability increased by, at most, a few percent for satellites close to the detection threshold.

Our catalog-level insertion procedure does not account for effects of blending in regions of high object density that might affect the detection and/or photometric measurements of member stars.
However, the constraints that we placed on the number of bright member stars and surface brightness typically limit our simulated satellite population to surface densities below a few stars per square arcminute (\appref{mu_ng22}).
Based on studies of the performance of the DESDM pipeline in crowded regions, blending will not substantially decrease the detectability of satellite galaxies with these surface densities \citep{Wang:2019}.
In addition, diffuse light from unresolved stars is a subdominant component of the flux for resolved systems at these distances.
These assumption are violated for bright nearby globular clusters and classical dwarf galaxies, but we assert that searches are complete for such systems in our survey area.

\begin{deluxetable}{l c c c }
\tabletypesize{\footnotesize}
\tablecaption{\label{tab:sim_params} Parameter ranges of simulated satellites}
\tablehead{\colhead{Parameter} & \colhead{Range} & \colhead{Unit} & Sampling }
\startdata
Stellar Mass &  $[10,\ 10^6]$ & \Msun & log \\
Heliocentric Distance & $[5,\ 10^3]$ & \kpc & log \\
2D Half-light Radius & $[1,\ 2\times10^3]$ & \pc & log \\
Ellipticity  & $[0.1,\ 0.8]$ & \ldots & linear \\
Position Angle & $[0,\ 180]$ & $\deg$ & linear \\
Age & $\{10,\ 12,\ 13.5\}$ & \Gyr & choice \\
Metallicity & $\{0.0001,\ 0.0002\}$ & \ldots & choice \\
\enddata
{\footnotesize \tablecomments{Simulated satellite properties are drawn from uniform distributions in log space with the parameter ranges listed above. Isochrones were generated using the models of \citet{Bressan:2012} with an IMF from \citet{Chabrier:2001}.}}
\end{deluxetable}

\section{Search Algorithms}\label{sec:algorithms}

Milky Way satellites are detected as arcminute-scale over-densities of old, metal poor stars located in the outer halo of the Milky Way. 
The brightest satellites were predominantly discovered in visual searches of photographic plates \citep{1938BHarO.908....1S,1938Natur.142..715S,1950PASP...62..118H,1955PASP...67...27W,1977MNRAS.180P..81C,1990MNRAS.244P..16I,1994Natur.370..194I}.
The advent of large digital sky surveys enabled the discovery of fainter systems using statistical matched-filter techniques \citep{2005AJ....129.2692W,2005ApJ...626L..85W,2006ApJ...650L..41Z,2006ApJ...643L.103Z,2006ApJ...647L.111B,2007ApJ...654..897B,2008ApJ...686L..83B,2009MNRAS.397.1748B,2010ApJ...712L.103B,2006ApJ...645L..37G,Grillmair:2009,2006ApJ...653L..29S,2007ApJ...656L..13I,2007ApJ...662L..83W}.
Matched-filter searches have been broadly applied to the current generation of large surveys to detect larger, fainter, and more distant systems \citep{Koposov:2015,Koposov:2018,Kim:2015d,Kim:2015b,Kim:2015c,martin_2015_hydra_ii,Laevens:2015a,Laevens:2015b,Torrealba:2016a,Torrealba:2016b,Torrealba:2018a,Torrealba:2019,Homma:2016,Homma:2017,Homma:2019,Luque:2017,Mau:2019b}.
In addition, maximum-likelihood-based algorithms have been developed to simultaneously combine morphological and photometric information to increase sensitivity \citep{Bechtol:2015,Drlica-Wagner:2015,Drlica-Wagner:2016hwk}.

Our search employed two automated algorithms to detect low-surface-brightness, arcminute-scale stellar overdensities. 
The first algorithm uses a conventional matched-filter approach, while the second uses a more complex maximum-likelihood framework.
Both search algorithms were optimized to detect old, metal-poor stellar populations using their distinct locus in color--magnitude space.
Our search focused on conventional ultra-faint galaxies and has slightly reduced efficiency for very large stellar systems \citep[\eg,][]{Torrealba:2016a,Pieres:2017,Torrealba:2019}, or especially young and/or metal rich systems \citep[\eg,][]{Torrealba:2018b}.
Our search was also optimized for high Galactic latitude, where the foreground stellar density does not vary significantly over degree scales. 
Importantly, the our two search methods employ different strategies to evaluate the local stellar density and to filter candidate member stars of Milky Way satellites according to their spatial and color measurements. 

\subsection{Spatial Matched-filter Search}
\label{sec:simple}

The first search algorithm, \simple, is inspired by the matched-filter methods of \citet{Koposov:2008} and \citet{Walsh:2009}, and uses a simple isochrone filter to enhance the contrast of halo substructures at a given distance relative to the foreground field of Milky Way stars.
The specific implementation builds upon the technique described by \citet{Bechtol:2015} and \citet{Drlica-Wagner:2015}.\footnote{\href{https://github.com/DarkEnergySurvey/simple}{https://github.com/DarkEnergySurvey/simple}}
When analyzing DES Y3A2, we required that objects be detected in both $g$ and $r$ bands and be brighter than $g = 24.5 \magn$.
When analyzing PS1 DR1, we adopted a signal-to-noise threshold $\rm{S/N} > 10$ in the $r$ band.
A matched-filter search for spatial overdensities of old, metal-poor stars was performed, scanning in distance modulus from $16 \leq \modulus \leq 24 \magn$ ($16 \leq \modulus \leq 22 \magn$) for DES Y3A2 (PS1 DR1) in steps of 0.5 \magn.
These searches correspond to heliocentric distances of $16 \kpc \leq D \leq 620 \kpc$ and $16\kpc \leq D \leq 251\kpc$, respectively.\footnote{Our search is less sensitive to systems at larger distances where the apparent magnitude of horizontal branch stars is fainter than the detection limit of the surveys.} 
At each distance modulus, we selected stars with $g$- and $r$-band magnitudes consistent with the synthetic isochrone of \citet{Bressan:2012} with metallicity $\metal = 0.0001$ and age $\age = 12\Gyr$. 
We required that the color difference between each star and the template isochrone be $\Delta (g-r) < \sqrt{0.1^2 + \sigma_g^2 + \sigma_r^2}$, where $\sigma_g$ and $\sigma_r$ are the statistical uncertainties on the $g$- and $r$-band magnitudes, respectively.

The survey footprint was partitioned into \healpix pixels of $\nside=32$ ($\roughly 3.4 \deg^2$) for individual analysis.
For each $\nside=32$ pixel and distance modulus step, we applied the isochrone filter described previously and created a map of the filtered stellar density field, including the central pixel of interest along with the eight surrounding \healpix pixels.
The eight surrounding pixels were used to more accurately estimate the average stellar density in the central pixel of interest.
The filtered stellar density field in the central pixel was smoothed by a Gaussian kernel ($\sigma = 2\arcmin$), and we identified local density peaks by iteratively raising a density threshold until there are fewer than 10 disconnected regions above the threshold value.
In practice, only the most prominent of these stellar overdensities passed our minimal statistical significance thresholds.

At the central location of each density peak, we determined the angular size of a surrounding aperture that maximizes the significance of the density peak with respect to the distribution of field stars. 
Specifically, we iterate through circular apertures with radii from $1\arcmin$ to $18\arcmin$, and for each radius, we compute the Poisson significance for the observed stellar counts within the aperture given the local field density.
The local field density is estimated from an annulus between $18\arcmin$ and $30\arcmin$ surrounding the peak. 
When calculating the stellar density, we account for the coverage of the survey, which is mapped at square arcminute scales, as described in \secrefs{des}{ps1}.
After consolidating spatially coincident peaks at different distance moduli, all peaks with Poisson significance ${\rm SIG} > 5.5 \sigma$ are considered seeds for subsequent analysis.
\simple has a high significance ceiling at ${\rm SIG} = 37.5 \sigma$, corresponding to the numerical limit of the inverse survival function of the normal distribution implemented in \code{scipy}.

\subsection{Likelihood-based Search}
\label{sec:ugali}

The second search algorithm employs a likelihood-based approach implemented with the \ugali framework \citep{Bechtol:2015,Drlica-Wagner:2015}.\footnote{\href{https://github.com/DarkEnergySurvey/ugali}{https://github.com/DarkEnergySurvey/ugali}}
A likelihood function is constructed from the product of Poisson probabilities to detect individual stars based upon their spatial positions, measured fluxes, photometric uncertainties, and the local imaging depth, given a model that includes a putative dwarf galaxy and empirical estimation of the local stellar field population (see \appref{like} for more details).
When calculating the likelihood, we account for both missing survey area and local depth variations mapped on square-arcminute scales (\secrefs{des}{ps1}).
We assumed a radially symmetric Plummer profile, scanning over half-light radii, $r_h = \{1\farcm2, 4\farcm2, 9\farcm0\}$, and a spectral model composed of four \citet{Bressan:2012} isochrones of \age = \{10 Gyr, 12 Gyr\} and \metal = \{0.0001, 0.0002\}, each weighted by a \citet{Chabrier:2001} IMF.
This spatial-spectral template was rastered over a spatial grid of \healpix pixels (\nside = 4096; spatial resolution of $\roughly 0\farcm7$) and range of distance moduli from $16 < \modulus < 23$ (heliocentric distances of $16 \kpc < D < 400 \kpc$) in steps of 0.5 \magn. 
At each coordinate, we evaluated the likelihood ratio between models with and without a candidate satellite galaxy to generate a three-dimensional map of detection significance.
We define a test statistic, $\TS = -2 \Delta \log(\mathcal{L})$, as our criterion for detection.
In the asymptotic limit, the \TS will follow a $\chi^2$-distribution with $n$ degrees of freedom \citep{Wilks:1938, Chernoff:1954}.
In our case, the grid scan maximizes over a grid of satellite sky location, distance, richness, and size, yielding $n \sim 5$, and our threshold of $\sqrt{\TS} > 6$ corresponds to a statistical significance of $\roughly 4.9\sigma$.
Isolated peaks in the \TS map were extracted as seeds for further characterization.

\startlongtable
\newcommand{\satscaption}{Confirmed and candidate Milky Way satellites\label{tab:all_mw_sats}}
\newcommand{\satscomments}{ 
Column descriptions: (1) satellite name, (2) survey(s) in which the system resides, (3) system classification (see below), (4, 5) published right ascension and declination, (6, 7, 8) published distance modulus, observed semi-major axis of an ellipse containing half of the light, and ellipticity, (9, 10) derived heliocentric distance and azimuthally averaged physical half-light radius, (11) published absolute $V$-band magnitude, (12) literature reference. 
When two references are listed, the second was used for the distance measurement.
Classifications are: (1) unconfirmed systems, (2) probable star clusters, (3) probable dwarfs, (4) kinematically confirmed dwarfs.
}
\newcommand{\satsnotes}{
\tablenotetext{a}{Approximate half-light radius and ellipticity estimated from \citet{Grillmair:2009}. }
}
\begin{deluxetable*}{lcc @{\hspace{0.3in}} ccccc @{\hspace{0.3in}} ccc @{\hspace{0.3in}} c}
\tabletypesize{\scriptsize}
\tablewidth{0pc}
\tablecolumns{12}
\tablecaption{\satscaption}
\tablehead{
\colhead{(1)} & \colhead{(2)} & \colhead{(3)}\hspace{0.25in} & \colhead{(4)} & \colhead{(5)} & \colhead{(6)} & \colhead{(7)} & \colhead{(8)}\hspace{0.25in} & \colhead{(9)} & \colhead{(10)} & \colhead{(11)}\hspace{0.25in} & \colhead{(12)} \\[-0.5em]
 \colhead{Name} & \colhead{Survey} & \colhead{Classification}\hspace{0.25in} & \colhead{RA} & \colhead{Dec} & \colhead{$m - M$} & \colhead{$a_h$} & \colhead{$\epsilon$}\hspace{0.25in} & \colhead{Distance} & \colhead{$r_{1/2}$} & \colhead{$M_V$}\hspace{0.25in} & \colhead{Ref.} \\[-0.5em]
 \colhead{} & \colhead{} & \colhead{}\hspace{0.25in} & \colhead{($\deg$)} & \colhead{($\deg$)} & \colhead{} & \colhead{($'$)} & \colhead{}\hspace{0.25in} & \colhead{(kpc)} & \colhead{(pc)} & \colhead{(mag)}\hspace{0.25in} & \colhead{}  }
\startdata
Antlia II &  & 4 & 143.8868 & -36.7673 & 20.6 & 76.2 & 0.38 & 132 & 2301 & -9.03 & 1\\
Aquarius II & PS1 & 4 & 338.4813 & -9.3274 & 20.2 & 5.1 & 0.39 & 108 & 125 & -4.4 & 2\\
Bo{\"o}tes I & PS1 & 4 & 210.0200 & 14.5135 & 19.1 & 9.97 & 0.30 & 66 & 160 & -6.02 & 3\\
Bo{\"o}tes II & PS1 & 4 & 209.5141 & 12.8553 & 18.1 & 3.17 & 0.25 & 42 & 33 & -2.94 & 3\\
Bo{\"o}tes III\tablenotemark{a} & PS1 & 4 & 209.3 & 26.8 & 18.4 & 30.0 & 0.5 & 47 & 289 & -5.75 & 4\\
Bo{\"o}tes IV & PS1 & 3 & 233.689 & 43.726 & 21.6 & 7.6 & 0.64 & 209 & 277 & -4.53 & 5\\
Canes Venatici I & PS1 & 4 & 202.0091 & 33.5521 & 21.7 & 7.12 & 0.44 & 218 & 338 & -8.80 & 3\\
Canes Venatici II & PS1 & 4 & 194.2927 & 34.3226 & 21.0 & 1.52 & 0.40 & 160 & 55 & -5.17 & 3\\
Carina &  & 4 & 100.4065 & -50.9593 & 20.1 & 10.1 & 0.36 & 105 & 248 & -9.43 & 3\\
Carina II &  & 4 & 114.1066 & -57.9991 & 17.8 & 8.69 & 0.34 & 36 & 77 & -4.5 & 6\\
Carina III &  & 4 & 114.6298 & -57.8997 & 17.2 & 3.75 & 0.55 & 28 & 20 & -2.4 & 6\\
Centaurus I &  & 3 & 189.585 & -40.902 & 20.3 & 2.9 & 0.4 & 116 & 76 & -5.55 & 7\\
Cetus II & PS1, DES & 3 & 19.47 & -17.42 & 17.4 & 1.9 & $< 0.4$ & 30 & 17 & 0.0 & 8\\
Cetus III & PS1, DES & 3 & 31.331 & -4.270 & 22.0 & 1.23 & 0.76 & 251 & 44 & -2.5 & 9\\
Columba I & PS1, DES & 3 & 82.86 & -28.01 & 21.3 & 2.2 & 0.3 & 183 & 98 & -4.2 & 10\\
Coma Berenices & PS1 & 4 & 186.7454 & 23.9069 & 18.2 & 5.64 & 0.37 & 44 & 57 & -4.38 & 3\\
Crater II & PS1 & 4 & 177.310 & -18.413 & 20.4 & 31.2 & $< 0.1$ & 117 & 1066 & -8.2 & 11\\
DES J0225+0304 & DES & 1 & 36.4267 & 3.0695 & 16.9 & 2.68 & 0.61 & 24 & 12 & -1.1 & 12\\
Draco & PS1 & 4 & 260.0684 & 57.9185 & 19.4 & 9.67 & 0.29 & 76 & 180 & -8.71 & 3\\
Draco II & PS1 & 3 & 238.174 & 64.579 & 16.7 & 3.0 & 0.23 & 22 & 17 & -0.8 & 13\\
Eridanus II & DES & 4 & 56.0925 & -43.5329 & 22.9 & 1.77 & 0.35 & 380 & 158 & -7.21 & 3\\
Fornax & DES & 4 & 39.9583 & -34.4997 & 20.8 & 19.6 & 0.29 & 147 & 707 & -13.46 & 3, 14\\
Grus I & DES & 3 & 344.1797 & -50.18 & 20.4 & 0.81 & 0.45 & 120 & 21 & -3.47 & 3\\
Grus II & DES & 3 & 331.02 & -46.44 & 18.6 & 6.0 & $< 0.2$ & 53 & 92 & -3.9 & 8\\
Hercules & PS1 & 4 & 247.7722 & 12.7852 & 20.6 & 5.63 & 0.69 & 132 & 120 & -5.83 & 3\\
Horologium I & DES & 4 & 43.8813 & -54.116 & 19.5 & 1.59 & 0.27 & 79 & 31 & -3.55 & 3\\
Horologium II & DES & 3 & 49.1077 & -50.0486 & 19.5 & 2.09 & 0.52 & 78 & 33 & -2.6 & 15\\
Hydra II &  & 3 & 185.4251 & -31.9860 & 20.9 & 1.52 & 0.24 & 151 & 58 & -4.60 & 3\\
Hydrus I &  & 4 & 37.389 & -79.3089 & 17.2 & 7.42 & 0.21 & 28 & 53 & -4.71 & 16\\
Indus II & DES & 1 & 309.72 & -46.16 & 21.7 & 2.9 & $< 0.4$ & 214 & 180 & -4.3 & 8\\
Kim 2 & DES & 2 & 317.2020 & -51.1671 & 20.0 & 0.48 & 0.32 & 100 & 12 & -3.32 & 3\\
Laevens 1 & PS1 & 2 & 174.0668 & -10.8772 & 20.8 & 0.51 & 0.17 & 145 & 20 & -4.80 & 3\\
LMC &  & 4 & 80.8938 & -69.7561 & 18.5 & 323.0 & 0.15 & 50 & 4735 & -18.12 & 17, 18\\
Leo I & PS1 & 4 & 152.1146 & 12.3059 & 22.0 & 3.65 & 0.3 & 254 & 226 & -11.78 & 3\\
Leo II & PS1 & 4 & 168.3627 & 22.1529 & 21.8 & 2.52 & 0.07 & 233 & 165 & -9.74 & 3\\
Leo IV & PS1 & 4 & 173.2405 & -0.5453 & 20.9 & 2.54 & 0.17 & 154 & 104 & -4.99 & 3\\
Leo V & PS1 & 4 & 172.7857 & 2.2194 & 21.3 & 1.00 & 0.43 & 178 & 39 & -4.40 & 3\\
Pegasus III & PS1 & 4 & 336.102 & 5.405 & 21.7 & 0.85 & 0.38 & 215 & 42 & -3.4 & 19\\
Phoenix II & DES & 4 & 354.996 & -54.4115 & 19.6 & 1.49 & 0.67 & 83 & 21 & -3.30 & 3\\
Pictor I & DES & 3 & 70.949 & -50.2854 & 20.3 & 0.88 & 0.63 & 114 & 18 & -3.45 & 3\\
Pictor II &  & 3 & 101.180 & -59.897 & 18.3 & 3.8 & 0.13 & 46 & 47 & -3.2 & 20\\
Pisces II & PS1 & 4 & 344.6345 & 5.9526 & 21.3 & 1.12 & 0.34 & 182 & 48 & -4.22 & 3\\
Reticulum II & DES & 4 & 53.9203 & -54.0513 & 17.4 & 5.52 & 0.58 & 30 & 31 & -3.88 & 3\\
Reticulum III & DES & 3 & 56.36 & -60.45 & 19.8 & 2.4 & $< 0.4$ & 92 & 64 & -3.3 & 8\\
Sagittarius &  & 4 & 283.8313 & -30.5453 & 17.1 & 342.0 & 0.64 & 26 & 1565 & -13.5 & 21\\
Sagittarius II & PS1 & 4 & 298.1647 & -22.0651 & 19.2 & 1.6 & $< 0.1$ & 69 & 32 & -5.2 & 22\\
Sculptor & DES & 4 & 15.0183 & -33.7186 & 19.6 & 11.17 & 0.33 & 84 & 223 & -10.82 & 3, 23\\
Segue 1 & PS1 & 4 & 151.7504 & 16.0756 & 16.8 & 3.62 & 0.33 & 23 & 20 & -1.30 & 3\\
Segue 2 & PS1 & 3 & 34.8226 & 20.1624 & 17.7 & 3.76 & 0.22 & 35 & 34 & -1.86 & 3\\
Sextans & PS1 & 4 & 153.2628 & -1.6133 & 19.7 & 16.5 & 0.30 & 86 & 345 & -8.72 & 3\\
SMC &  & 4 & 13.1867 & -72.8286 & 19.0 & 151.0 & 0.40 & 62 & 2728 & -17.18 & 17, 24\\
Triangulum II & PS1 & 3 & 33.3252 & 36.1702 & 17.4 & 1.99 & 0.46 & 30 & 13 & -1.60 & 3\\
Tucana II & DES & 4 & 342.9796 & -58.5689 & 18.8 & 12.59 & 0.39 & 58 & 165 & -3.8 & 25\\
Tucana III & DES & 3 & 359.15 & -59.60 & 17.0 & 6.0 & 0.0 & 25 & 44 & -2.4 & 8\\
Tucana IV & DES & 4 & 0.73 & -60.85 & 18.4 & 11.8 & 0.4 & 48 & 128 & -3.5 & 8\\
Tucana V & DES & 3 & 354.35 & -63.27 & 18.7 & 1.8 & 0.7 & 55 & 16 & -1.6 & 8\\
Ursa Major I & PS1 & 4 & 158.7706 & 51.9479 & 19.9 & 8.31 & 0.59 & 97 & 151 & -5.12 & 3\\
Ursa Major II & PS1 & 4 & 132.8726 & 63.1335 & 17.5 & 13.8 & 0.56 & 32 & 85 & -4.25 & 3\\
Ursa Minor & PS1 & 4 & 227.2420 & 67.2221 & 19.4 & 18.3 & 0.55 & 76 & 272 & -9.03 & 3\\
Virgo I & PS1 & 3 & 180.038 & -0.681 & 19.8 & 1.76 & 0.59 & 91 & 30 & -0.33 & 9\\
Willman 1 & PS1 & 4 & 162.3436 & 51.0501 & 17.9 & 2.51 & 0.47 & 38 & 20 & -2.53 & 3\\
\enddata
{\footnotesize \tablecomments{\satscomments Literature references are: (1) \cite{Torrealba:2019}, (2) \cite{Torrealba:2016b}, (3) \cite{Munoz:2018}, (4) \cite{Grillmair:2009}, (5) \cite{Homma:2019}, (6) \cite{Torrealba:2018a}, (7) \cite{Mau:2019b}, (8) \cite{Drlica-Wagner:2015}, (9) \cite{Homma:2017}, (10) \cite{Carlin:2017}, (11) \cite{Torrealba:2016a}, (12) \cite{Luque:2017}, (13) \cite{Longeard:2019}, (14) \cite{Pietrzynski:2009}, (15) \cite{Kim:2015c}, (16) \cite{Koposov:2018}, (17) \cite{Makarov:2014}, (18) \cite{Pietrzynski:2013}, (19) \cite{Kim:2016}, (20) \cite{Drlica-Wagner:2016hwk}, (21) \cite{McConnachie:2012}, (22) \cite{Mutlu-Pakdil:2018}, (23) \cite{Martinez:2015}, (24) \cite{Graczyk:2014}, (25) \cite{Koposov:2015}.}}
\satsnotes
\end{deluxetable*}

\section{Search Results}\label{sec:results}

The DES and PS1 searches each returned several thousand ``seeds'' (\ie, locations where the local significance exceeds a minimum threshold).
The observed distribution of detection significance falls steeply for each algorithm (\figref{mask_hists}), with the majority of high-significance seeds coinciding with real resolved stellar systems, regions of spatially varying Galactic extinction or stellar density, or survey artifacts caused by bright stars and incomplete coverage.
To robustly infer the population of Milky Way satellites, we define additional criteria to produce a high-purity sample of ``detected'' satellite candidates.
These criteria were self-consistently applied to both survey data and simulations.

\subsection{Detection Criteria}
\label{sec:detect}

\begin{figure*}
\includegraphics[width=1.\textwidth]{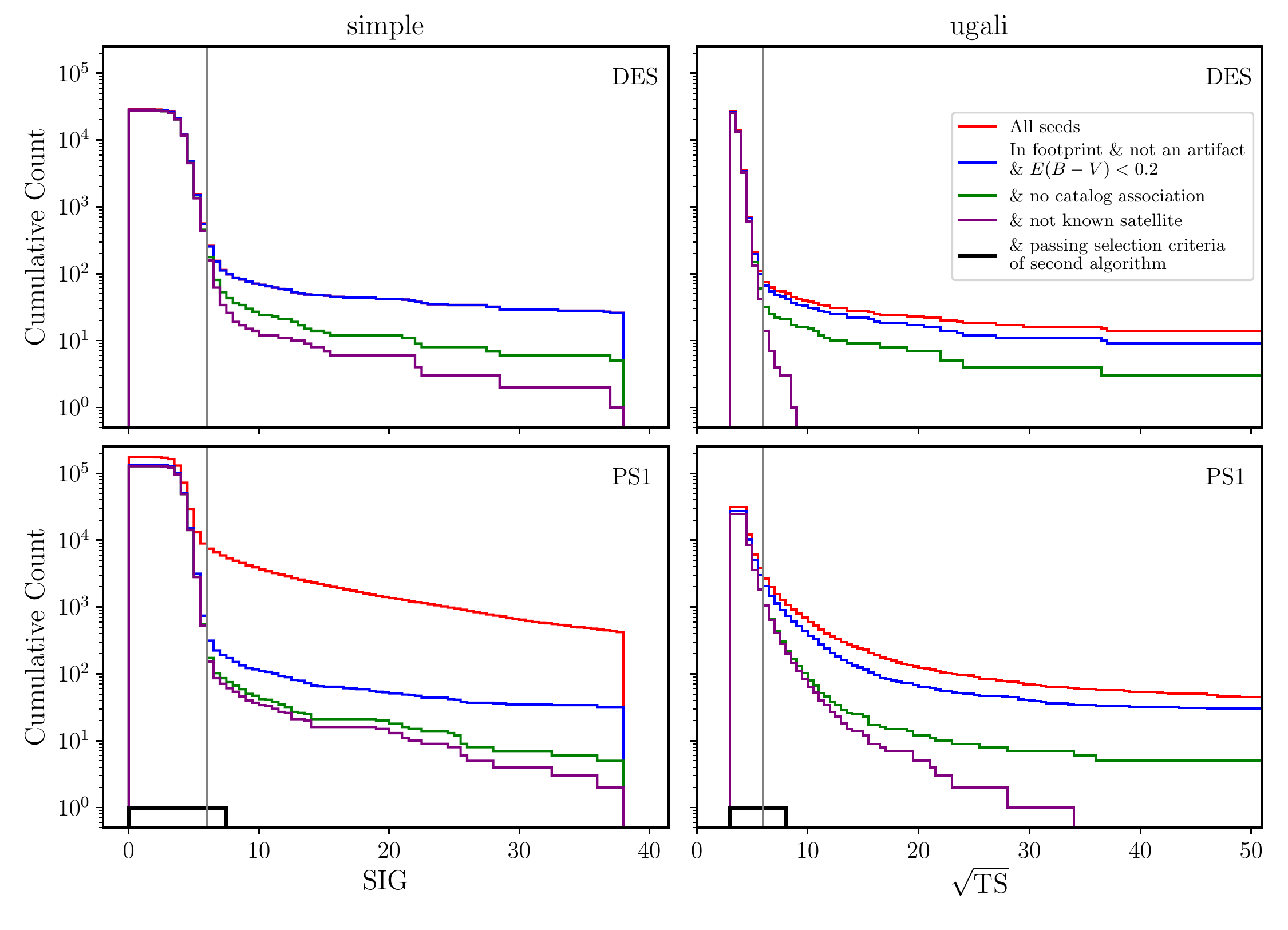}
\caption{Distribution of detection significance for seeds returned by the \simple and \ugali search algorithms in our DES and PS1 satellite searches. 
Histograms show seeds that survive successive application of selection criteria, including the geometric masks shown in Figure \ref{fig:masks}.
The solid vertical line denotes our fiducial detection significance threshold for each survey. Note that the \simple algorithm returns a maximum significance of 37.5. \KB{The ``detected by both algorithms'' curve is different from the others. For example, in DES \ugali panel, what we really mean is ``detected by \simple above benchmark threshold'', and in the DES \simple panel, we mean ``detected by \ugali above benchmark threshold''. In other words, in the \ugali case, we scan in $\sqrt{\rm TS}$ while holding \SIG fixed.}}
\label{fig:mask_hists}
\end{figure*}

We developed a set of detection criteria intended to refine the set of raw seeds to a list of high-quality satellite galaxy candidates.
These criteria were partially motivated by the observed distribution of seeds and the recovery of previously known satellites.
We sought to design a set of selection criteria for which our recovery of known satellites was relatively complete, but restrictive enough that only a small number of additional objects would pass our criteria and could be visually inspected. 

Our detection criteria can be broadly categorized into four different types:
(i) a set of geometric criteria intended to mask known stellar systems and problematic regions of the survey (\figref{masks});
(ii) a detection significance threshold; 
(iii) a spatial match between seeds from the \simple and \ugali searches; and
(iv) a visual inspection and masking of residual survey artifacts.
These criteria mimic the criteria applied for satellite discovery in previous studies \citep[\ie,][]{Bechtol:2015,Drlica-Wagner:2015} and are described in more detail below.

\paragraph{(i) Geometric Criteria} We applied a set of geometric masks to exclude regions of the survey footprint where systematic features result in the detection of spurious seeds.
We began by restricting the seeds from both surveys to regions of low interstellar extinction, $E(B-V) < 0.2$ \citep{Schlegel:1998}. 
While our algorithms incorporate the effects of reddening, regions of high reddening generally trace regions of high Milky Way stellar density. 
In these regions, the reddening and the stellar field density can vary over relatively small spatial scales, and we find the incidence of false-positive seeds is disproportionately high.
This mask removes $\roughly 7400 \deg^2$ from the PS1 footprint and is negligible for DES (\figref{masks}).

Our empirical model of the PS1 DR1 footprint is inaccurate near the survey boundaries, resulting in the detection of spurious seeds due to mis-estimation of the stellar density. 
To remove these spurious seeds, we applied a declination selection of $\delta > -25\deg$. 
We also removed regions of the PS1 footprint where very high stellar density led to memory overflow issues during application of the likelihood search algorithm.
These regions largely overlap with the reddening mask and only remove $\roughly 20\deg^2$ of additional area.

We also masked regions around astronomical objects that are known to produce spurious seeds.
These masks can generally be separated into regions around bright stars \citep{1991bsc..book.....H}, Milky Way globular clusters \citep[][2010 edition]{Harris96}, open clusters (WEBDA)\footnote{\url{https://webda.physics.muni.cz}}, and nearby galaxies that are resolved into individual stars \citep{2004yCat.7239....0H, 1973ugcg.book.....N, 1985IAUS..113..541W, 2013A&A...558A..53K, 2008MNRAS.389..678B}.
We also mask regions around overdensities in two narrow stellar streams, ATLAS \citep{Koposov:2014,Shipp:2018} and Phoenix \citep{Balbinot:2016}.
For extended objects, we masked regions consistent with the half-light radii of these objects, with a minimum masked radius of $0.05 \deg$.
For bright stars and objects where size information is unavailable, we masked a circular region with a $0.1 \deg$ radius.

\paragraph{(ii) Significance Threshold} We require a fiducial significance threshold of $\SIG > 6$ for the \simple search algorithm and $\sqrt{\TS} > 6$ for the \ugali search algorithm. 
These significance thresholds were chosen such that the observed number of unassociated seeds increases rapidly if the threshold is reduced (\figref{mask_hists}).
In addition, most of the seeds above these thresholds can be readily classified as either genuine stellar systems or obvious survey artifacts (\appref{candidates}), whereas seeds below these thresholds are more often ambiguous.

\paragraph{(iii) Detection by Both Algorithms}
We required that seeds be detected above the stated significance thresholds by both the \simple and \ugali search algorithms.
To apply this criteria, we matched seeds between the two searches.
We defined two seeds to be matched if their centroids were within $12\farcm0$ ($0.2 \deg$).
We reiterate that our objective is to derive a statistically rigorous observational selection function that yields a high-purity list of candidates.
Requiring detection by both algorithms greatly increases the purity of our candidate list at a moderate cost to detection efficiency (\figref{mask_hists}).
We find that achieving similar purity with a single algorithm would require a higher detection threshold and would result in lower overall efficiency.
Requiring detection by both algorithms does result in the rejection of several known satellites in the PS1 footprint that were only detected by \ugali (i.e., Aquarius II, Columba I, Leo IV, Leo V, Pisces II, Ursa Major I) or \simple (Bo\"{o}tes II).
We discuss these specific cases in \secref{recover_known}.

\paragraph{(iv) Residual Survey Artifacts}
We identified several seeds in the PS1 DR1 footprint that passed the previous selection criteria but were clearly associated with imaging or processing artifacts (\eg, stray and scattered light around bright stars, abrupt changes in survey depth, inaccurate survey coverage map, PSF fitting failures).
These regions were visually identified and masked with a circular mask of radius of $\roughly 0.3$--$0.5 \deg$. 
An additional \CHECK{12} seeds in the PS1 DR1 passed our fiducial selection criteria and were visually inspected. 
These seeds showed a poorly defined stellar sequence in color--magnitude space and/or poorly defined spatial morphology. 
These regions likely result from a combination of less obvious survey artifacts (\ie, mis-estimation of the sky background, excess sensor noise, or low level scattered light) and contamination from background galaxy clusters.
We mask regions of $\roughly 0.3 \deg$ around each of these seeds.
More details on the identification of these regions can be found in \appref{candidates}.

\medskip

The total search area after masking is \CHECK{$4{,}844 \deg^2$} in DES, \CHECK{$21{,}123 \deg^2$} in PS1, and \CHECK{$24{,}343 \deg^2$} together (DES and PS1 overlap in a region of $\roughly 1{,}600 \deg^2$).
\figref{mask_hists} shows the effect of the successive selections and demonstrates that our two independent search algorithms yield reasonably consistent results, particularly for objects that pass the selection criteria described above. 
The consistency in the number of high-significance objects returned by \emph{both} search algorithms lends confidence to our final list of satellite systems.

\subsection{Recovery of Real Satellites}
\label{sec:recover_known}

\newcommand{\knowncaptiondes}{Recovery of confirmed and candidate Milky Way satellite galaxies in DES Y3A2.\label{tab:known_des}}
\newcommand{\knowncommentsdes}{ 
Column descriptions: (1) satellite name; (2) square-root of the test statistic from \ugali search; (3) statistical significance value from \simple search (maximum of $37.5$); (4) detection probability from survey selection function; (5, 6) heliocentric distance and azimuthally averaged physical half-light radius, calculated from observed parameters listed in Table \ref{tab:all_mw_sats}; (7) published absolute $V$-band magnitude; (8) local stellar density. Satellites denoted with asterisks are not included in the statistical sample used to derive the luminosity function. 
}
\newcommand{\knownnotesdes}{}
\begin{deluxetable*}{lccccccc}
\tabletypesize{\scriptsize}
\tablewidth{0pc}
\tablecolumns{8}
\tablecaption{\knowncaptiondes}
\tablehead{
\colhead{(1)} & \colhead{(2)} & \colhead{(3)} & \colhead{(4)} & \colhead{(5)} & \colhead{(6)} & \colhead{(7)} & \colhead{(8)} \\[-0.5em]
 \colhead{Name} & \colhead{$\sqrt{\mathrm{TS}}$} & \colhead{SIG} & \colhead{$P_{\rm det}$} & \colhead{Distance} & \colhead{$r_{1/2}$} & \colhead{$M_V$} & \colhead{$\rho_\star$} \\[-0.5em]
 \colhead{} & \colhead{} & \colhead{} & \colhead{} & \colhead{(kpc)} & \colhead{(pc)} & \colhead{(mag)} & \colhead{(arcmin$^{-2}$)}  }
\startdata
Fornax & 480.07 & 37.5 & 1.00 & 147 & 707 & -13.46 & 16.52\\
Sculptor & 415.08 & 37.5 & 1.00 & 84 & 223 & -10.82 & 5.87\\
Reticulum II & 54.56 & 37.5 & 1.00 & 30 & 31 & -3.88 & 1.17\\
\**Eridanus II & 36.17 & 27.41 & 1.00 & 380 & 158 & -7.21 & 1.10\\
Tucana II & 23.85 & 12.88 & 0.91 & 58 & 165 & -3.8 & 1.87\\
Grus II & 21.82 & 13.29 & 0.97 & 53 & 92 & -3.9 & 2.03\\
Horologium I & 21.78 & 20.68 & 0.99 & 79 & 31 & -3.55 & 1.24\\
Tucana III & 18.60 & 12.72 & 0.91 & 25 & 44 & -2.4 & 1.62\\
Tucana IV & 16.03 & 10.88 & 0.91 & 48 & 128 & -3.5 & 1.60\\
Phoenix II & 13.36 & 12.18 & 0.98 & 83 & 21 & -3.30 & 1.35\\
Horologium II & 11.62 & 11.21 & 0.87 & 78 & 33 & -2.6 & 1.07\\
Tucana V & 11.27 & 9.97 & 0.89 & 55 & 16 & -1.6 & 1.78\\
Pictor I & 10.91 & 8.55 & 0.97 & 114 & 18 & -3.45 & 1.50\\
Columba I & 10.67 & 9.45 & 0.33 & 183 & 98 & -4.2 & 2.09\\
Cetus II & 10.47 & 7.18 & 0.62 & 30 & 17 & 0.0 & 0.79\\
Grus I & 9.78 & 8.80 & 0.97 & 120 & 21 & -3.47 & 1.57\\
\**Kim 2 & 8.17 & 9.31 & 0.93 & 100 & 12 & -3.32 & 3.13\\
Reticulum III & 8.12 & 7.46 & 0.92 & 92 & 64 & -3.3 & 1.46\\
\**Cetus III & 3.96 & 3.68 & 0.01 & 251 & 44 & -2.5 & 0.89\\
\**Indus II & 3.86 & 3.65 & 0.01 & 214 & 180 & -4.3 & 4.02\\
\**DES J0225+0304 & \ldots & \ldots & 0.94 & 24 & 12 & -1.1 & 0.98\\
\enddata
{\footnotesize \tablecomments{\knowncommentsdes}}
\knownnotesdes
\end{deluxetable*}

\newcommand{\knowncaptionps}{Recovery of confirmed and candidate Milky Way satellite galaxies in PS1 DR1.\label{tab:known_ps1}}
\newcommand{\knowncommentsps}{
Column descriptions are the same as \tabref{known_des}. Satellites denoted with asterisks are not included in the statistical sample used to derive the luminosity function. 
}
\newcommand{\knownnotesps}{
\tablenotetext{a}{Located in a masked region of the PS1 footprint ($\delta < -25 \deg$).}
\tablenotetext{b}{Approximate half-light radius and ellipticity estimated from \citet{Grillmair:2009}. }
}
\begin{deluxetable*}{lccccccc}
\tabletypesize{\scriptsize}
\tablewidth{0pc}
\tablecolumns{8}
\tablecaption{\knowncaptionps}
\tablehead{
\colhead{(1)} & \colhead{(2)} & \colhead{(3)} & \colhead{(4)} & \colhead{(5)} & \colhead{(6)} & \colhead{(7)} & \colhead{(8)} \\[-0.5em]
 \colhead{Name} & \colhead{$\sqrt{\mathrm{TS}}$} & \colhead{SIG} & \colhead{$P_{\rm det}$} & \colhead{Distance} & \colhead{$r_{1/2}$} & \colhead{$M_V$} & \colhead{$\rho_\star$} \\[-0.5em]
 \colhead{} & \colhead{} & \colhead{} & \colhead{} & \colhead{(kpc)} & \colhead{(pc)} & \colhead{(mag)} & \colhead{(arcmin$^{-2}$)}  }
\startdata
Leo I & 157.63 & 37.5 & 1.00 & 254 & 226 & -11.78 & 1.18\\
Leo II & 104.05 & 37.5 & 1.00 & 233 & 165 & -9.74 & 3.09\\
Draco & 96.94 & 37.5 & 1.00 & 76 & 180 & -8.71 & 3.02\\
Ursa Minor & 83.14 & 37.5 & 1.00 & 76 & 272 & -9.03 & 3.19\\
Sextans & 58.62 & 24.63 & 1.00 & 86 & 345 & -8.72 & 3.00\\
Canes Venatici I & 36.00 & 25.33 & 1.00 & 218 & 338 & -8.80 & 1.01\\
Bo{\"o}tes I & 25.29 & 11.63 & 0.95 & 66 & 160 & -6.02 & 1.29\\
Ursa Major II & 18.66 & 8.86 & 0.94 & 32 & 85 & -4.25 & 1.76\\
Coma Berenices & 15.29 & 9.75 & 0.93 & 44 & 57 & -4.38 & 1.07\\
Sagittarius II & 15.19 & 11.66 & 0.55 & 69 & 32 & -5.2 & 12.60\\
Willman 1 & 15.03 & 12.54 & 0.54 & 38 & 20 & -2.53 & 0.95\\
Canes Venatici II & 11.70 & 8.78 & 0.93 & 160 & 55 & -5.17 & 0.98\\
Segue 1 & 10.79 & 8.55 & 0.48 & 23 & 20 & -1.30 & 1.22\\
Segue 2 & 10.75 & 7.25 & 0.05 & 35 & 34 & -1.86 & 1.55\\
Crater II & 10.42 & 6.08 & 0.06 & 117 & 1066 & -8.2 & 2.44\\
\**Ursa Major I & 10.18 & 5.99 & 0.24 & 97 & 151 & -5.12 & 0.99\\
\**Laevens 1 & 9.89 & 9.52 & 0.96 & 145 & 20 & -4.8 & 1.67\\
Draco II & 9.76 & 7.90 & 0.24 & 22 & 17 & -0.8 & 1.63\\
Triangulum II & 9.46 & 6.76 & 0.23 & 30 & 13 & -1.60 & 3.04\\
Hercules & 9.11 & 6.44 & 0.44 & 132 & 120 & -5.83 & 3.91\\
\**Leo IV & 8.25 & 4.94 & 0.18 & 154 & 104 & -4.99 & 1.48\\
Cetus II & 7.42 & 6.14 & 0.02 & 30 & 17 & 0.0 & 1.08\\
\**Aquarius II & 7.27 & 5.07 & 0.01 & 108 & 125 & -4.4 & 1.67\\
\**Leo V & 6.95 & 4.14 & 0.14 & 178 & 39 & -4.4 & 1.31\\
\**Pisces II & 6.25 & 4.39 & 0.03 & 182 & 48 & -4.22 & 1.58\\
\**Columba I\tablenotemark{a} & 6.15 & 5.34 & 0.00 & 183 & 98 & -4.2 & 2.46\\
\**Bo{\"o}tes II & 5.95 & 6.46 & 0.42 & 42 & 33 & -2.94 & 1.27\\
\**Bo{\"o}tes IV & 5.41 & 4.69 & 0.00 & 209 & 277 & -4.53 & 1.56\\
\**Pegasus III & 4.80 & 3.68 & 0.00 & 215 & 42 & -3.4 & 2.09\\
\**Virgo I & 4.06 & 4.08 & 0.00 & 91 & 30 & -0.33 & 1.58\\
\**Bo{\"o}tes III\tablenotemark{b} & 4.00 & 4.34 & 0.24 & 47 & 289 & -5.75 & 1.16\\
\**Cetus III & \ldots & 4.55 & 0.00 & 251 & 44 & -2.5 & 1.00\\
\enddata
{\footnotesize \tablecomments{\knowncommentsps}}
\knownnotesps
\end{deluxetable*}

We compare the results of our search to the population of confirmed and candidate dwarf galaxies (\tabref{all_mw_sats}).
To assemble our catalog of satellites, we augmented \cite{McConnachie:2012} with other recently discovered ultra-faint satellites collected in \cite{Simon:2019}.
Our structural parameters are taken primarily from \citet{Munoz:2018}, incorporating improved measurements from deeper data when available (as noted in the table).
The kinematic classification of ultra-faint satellite systems is notoriously difficult, due to their faintness and small intrinsic velocity dispersions.
We thus assemble our catalog from larger satellite systems ($r_{1/2} > 100 \pc$) or smaller satellites with low average surface brightness ($10\pc \leq r_{1/2} \leq 100 \pc$ and $\mu > 25 \magn \asec^{-2}$). 
Classification for systems with $10\pc < r_{1/2} < 20\pc$ is particularly challenging since velocity dispersions are rarely resolved and classification arguments are often based on non-uniform chemical and structural measurements.
We divide  systems in this table into kinematically classified dwarf galaxies (class 4); probable dwarf galaxies based on structural or metallicity measurements (class 3); probable star clusters based on structural, age, or metallicity arguments (class 2); and unconfirmed systems that were discovered in shallower DES data but were not detected in our search (class 1).
Class 2 notably includes Crater I/Laevens 1 \citep{2014MNRAS.441.2124B,Laevens:2014a} and Kim 2 \citep{Kim:2015b}, which have been proposed to be star clusters due to structural, age, and metallicity arguments \citep{Kirby:2015,Weisz:2016}, but pass our selection on size and surface brightness.
Our primary sample of probable and kinematically classified satellite galaxies are systems with class $\geq3$.
Our search recovers the majority of previously discovered Milky Way satellite galaxies in the PS1 DR1 and DES Y3A2 footprints (\tabrefs{known_des}{known_ps1}).

We recovered \CHECK{18} out of \CHECK{21} confirmed and candidate satellite galaxies in the DES footprint above our nominal threshold of $\sqrt{\TS} > 6$ and $\SIG > 6$. 
Two of the ultra-faint galaxy candidates initially detected in DES Y2Q1 data were marked as ``lower-confidence'' candidates due to their locations in regions of non-uniform survey coverage during the first two years of DES observations \citep{Drlica-Wagner:2015}.
With the deeper and more homogeneous imaging of the Y3A2 data set, Cetus~II (\cetII) is detected with $\sqrt{\TS} = 10.5$ and $\SIG = 7.2$. 
This statistical significance is comparable to other confirmed candidates (\eg, Columba I and Pictor I). 
Meanwhile, the second low-confidence candidate, Indus~II (DES J2038\allowbreak$-$4609), drops below our detection threshold.\footnote{Deep follow-up imaging with Magellan/Megacam supports the hypothesis that Indus~II (DES J2038\allowbreak$-$4609) is a chance alignment of stars \citep{Cantu:2019}.} 
We recovered the Eridanus III system reported in \citet{Bechtol:2015} and \citet{Koposov:2015} with high significance ($\sqrt{\TS} = 9.5$ and $\SIG = 9.0$); however, due to its small physical size \citep[$r_{1/2} = 5 \pc$,][]{Conn:2018}, we do not include it in our list of confirmed and candidate dwarf galaxies.
Neither of the two objects reported in \citet{Luque:2017} was detected as a significant seed in our automated search, and visual inspection of the DES Y3A2 data coincident with these candidates did not reveal any significant excesses. 
Cetus~III, an ultra-faint galaxy candidate identified in early data from HSC SSP \citep{Homma:2017}, is located within the DES footprint but falls below our detection threshold.
This is expected, given the large distance (251 \kpc) and low azimuthally averaged surface brightness ($\roughly 29.3~\rm{mag\ arcsec}^{-2}$) of this candidate.

In comparison, we recovered \CHECK{20} of the \CHECK{32} confirmed and candidate satellite galaxies known to reside in the PS1 DR1 footprint.
The lower recovery rate in PS1 is expected since many of the satellites in the PS1 footprint were discovered with significantly deeper data.
Of the twelve satellites that fall below our detection threshold, five were discovered in deeper surveys and are not expected to be detected by PS1: Bo\"{o}tes IV, Cetus III, and Virgo I were discovered in HSC SSP; Columba I was discovered in DES; and Aquarius II was discovered in VST ATLAS. 
The seven remaining satellites were discovered using data from SDSS, but several of these objects were confirmed with deeper follow-up observations before publication.
Leo V had deep follow-up imaging from the Isaac Newton Telescope and spectroscopy from the Hectochelle fiber spectrograph at the Multiple Mirror Telescope \citep{2008ApJ...686L..83B}, while Pisces II had follow-up imaging from the MOSAIC camera at the 4-m Mayall Telescope \citep{2010ApJ...712L.103B}.
Pegasus III was announced after deep follow-up observations with DECam \citep{2015ApJ...799...73K}.
Leo IV was discovered in data from SDSS without additional follow-up \citep{2007ApJ...654..897B} and is detected significantly above threshold by \ugali ($\sqrt{\TS} = 8.2$). 
However, the \simple significance ($\SIG=4.9$) falls below our threshold.
Similarly, Ursa Major I is detected significantly with \ugali, but falls just below the threshold for \simple.
Bo\"{o}tes II falls just below our threshold for detection and has a comparably low detection probability reported by \cite{Koposov:2008}.
Bo\"{o}tes III is a diffuse object (extending $\roughly 1.5\deg$) with a complex morphology that was identified visually in filtered stellar density maps from SDSS DR5 \citep{Grillmair:2009}. 
The large size and complex morphology of Bo\"{o}tes III have made it challenging to detect with automated search algorithms \citep{Koposov:2008,Walsh:2009}.\footnote{We also examined the regions around five SDSS candidates proposed by \citet{Liu:2008}, but we do not find any significant excesses at these locations.}

We found one candidate in the PS1 DR1 search that passed all selection criteria and is unassociated with our catalogs of known satellites (see \appref{candidates} for more details).
This candidate, located at $(\ra,\dec) = (247.725,-0.971)$, is a compact, $r_{1/2} = 3.7\pc$, low-luminosity, $M_V = 0.6$, system residing at a heliocentric distance of $D = 15.6\kpc$. 
While the physical nature of such a faint system is ambiguous without internal kinematics measurements, the small physical size of this candidate is consistent with other low-luminosity outer-halo star clusters that have been discovered in recent surveys \citep[\eg,][]{Torrealba:2018b,Mau:2019}.
Due to the small physical size of this candidate, we classify it as a likely star cluster and do not include it in the sample of confirmed and candidates satellite galaxies used for deriving the Milky Way satellite galaxy luminosity function.

The non-detection of several known satellites is not unexpected, given that we prioritize purity over completeness in our selection criteria.
Discovery-driven searches often set a lower significance threshold, relying on visual inspection and targeted follow-up observations to reject false positives.
This subjective selection function is difficult to characterize with an automated analysis relying on simulations. 
By setting more restrictive detection criteria, we can be confident that every satellite that passes our automated selection criteria would pass subjective selection.
Such a requirement is critical to self-consistently interpret the recovery of simulated satellites and the derived observational selection function.

\subsection{Recovery of Simulated Satellites}
\label{sec:recover_sim}

\begin{figure*}
\centering
\includegraphics[width=\textwidth]{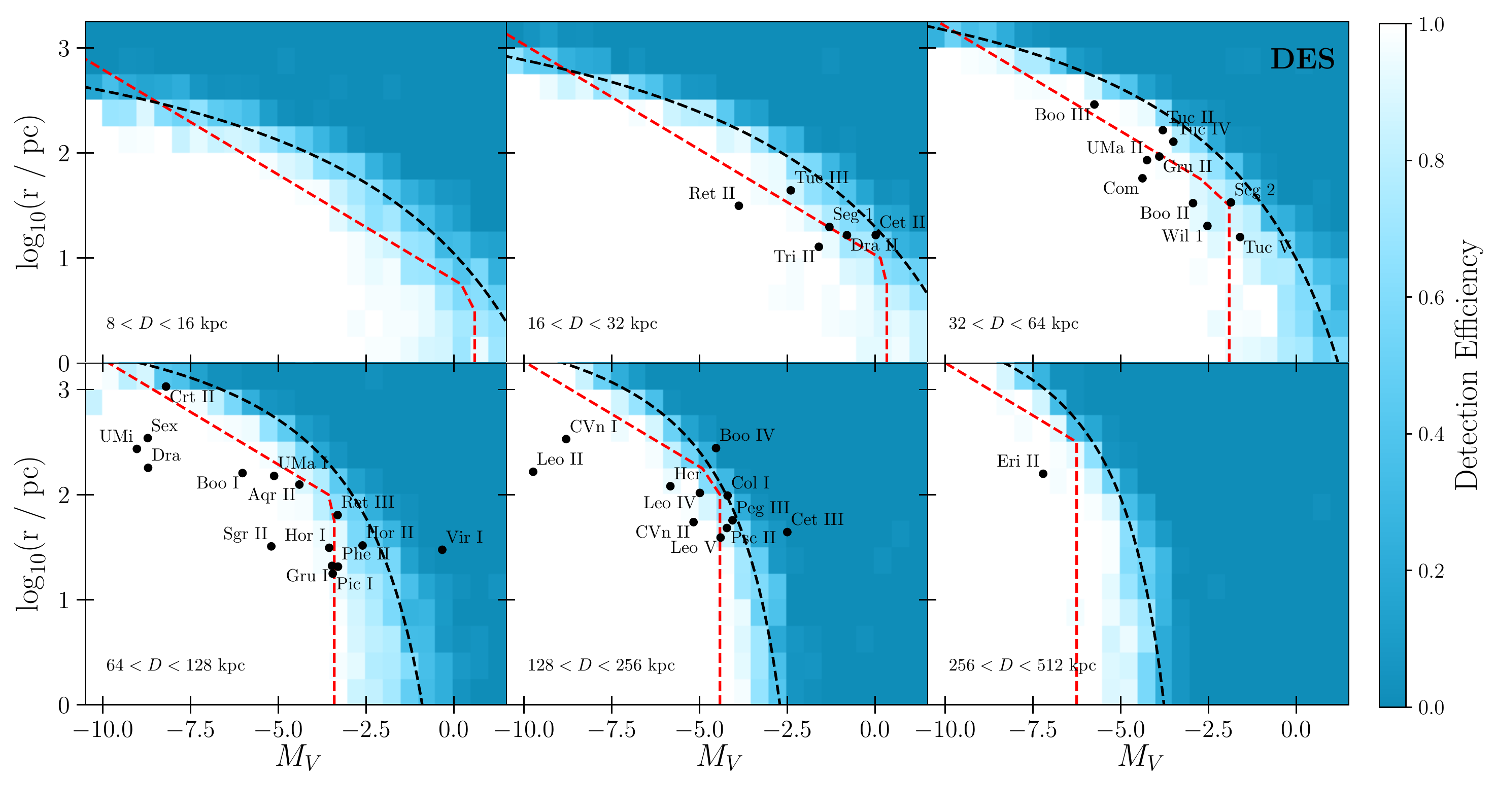}
\caption{\label{fig:des_slices} 
Detection efficiency of searches for Milky Way satellites in DES Y3A2.
Detection efficiency ranges from 0\% (blue) to 100\% (white) and is shown as a function of azimuthally averaged physical half-light radius and absolute $V$-band magnitude in different bins of heliocentric distance (logarithmically spaced from 8 kpc to 512 kpc).
The physical parameters of known satellites located within the DES and/or PS1 footprints are indicated in black. 
The black dashed line shows an analytic approximation to the 50\% detectability contour, while the red dashed line shows the 50\% detection efficiency contour for SDSS DR5 from \citet{Koposov:2008}.
The DES search is significantly more sensitive than the SDSS DR5 search of \citet{Koposov:2008}.
}
\end{figure*}

\begin{figure*}
\centering
\includegraphics[width=\textwidth]{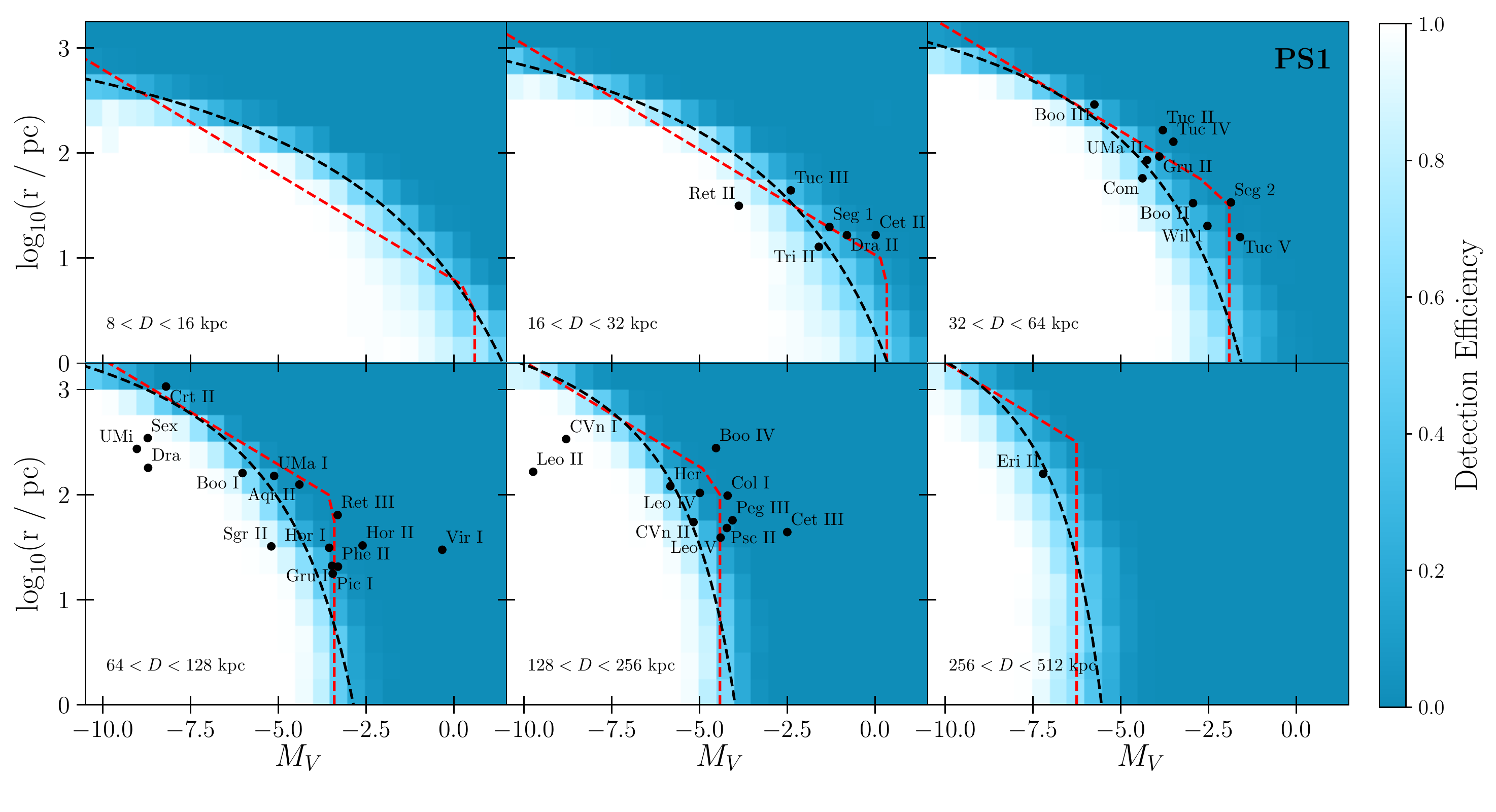}
\caption{\label{fig:ps1_slices} 
Similar to \figref{des_slices}, but for the PS1 DR1 search.
The 50\% detectability of the PS1 DR1 search (dashed black line) is significantly less sensitive than DES and is somewhat less sensitive than the 50\% detection efficiency of the SDSS DR5 search reported by \citet{Koposov:2008} (red dashed line).
}
\end{figure*}

We illustrate the recovery of simulated satellites in the physically motivated parameter space of satellite absolute magnitude, $M_V$, heliocentric distance, $D$, and azimuthally averaged physical half-light radius, $r_{1/2}$.
\figrefs{des_slices}{ps1_slices} show the detectability of simulated satellites as a function of $M_V$ and $r_{1/2}$ in six slices of distance. 
The coloring of each bin corresponds to the fraction of satellites in that bin that pass the detection criteria presented in \secref{detect} 
(\ie, $\sqrt{\TS} > 6$ and $\SIG > 6$).
The DES simulations generally contain $\roughly 30$ simulated satellites per bin, while the PS1 simulations contain $\roughly 250$ simulated satellites per bin.
We overplot the physical properties of the known Milky Way satellites residing within the DES and PS1 footprints from \tabrefs{known_des}{known_ps1}.
In general, the physical parameters of the known satellites recovered by our search (\secref{recover_known}) are consistent with the sensitivity envelope derived from simulated satellites (\figref{ps1_slices}).

We indicate the approximate surface brightness and absolute magnitude limits of SDSS as derived by \citet{Koposov:2008} with a dashed red line.
We find that our search on DES Y3A2 is significantly more sensitive than SDSS, while the sensitivity of our PS1 DR1 search is roughly comparable to the SDSS search in many regions of parameter space.
When directly comparing PS1 and SDSS catalogs in overlapping fields, and applying quality and star--galaxy separation criteria to both surveys, the stellar efficiency curves of SDSS and PS1 are similar.\footnote{Care must be taken in comparing the quoted depths of PS1 and SDSS.
SDSS conventionally quotes a 95\% completeness limit for point sources of $g,r = 22.2$ before star--galaxy separation. 
This is not directly comparable to the $10\sigma$ magnitude limit of $g,r \sim 22.5$ calculated for PS1 DR1 in \secref{data}.
The completeness depth of each data set is dependent on the exact selection criteria applied to the catalogs.
}
The sensitivity of our PS1 search is largely driven by the conservative detection thresholds we set and the requirement that candidates be detected by both search algorithms.
If these restrictions are relaxed, we find that our search becomes significantly more sensitive, but with a corresponding increase in the number of false-positive seeds that need to be rejected using visual inspection.
It may be possible to increase the sensitivity of the PS1 search if the incidence of spurious seeds can be reduced by subsequent PS1 data releases (\ie, PS1 DR2) or through a more precise estimate of the PS1 survey coverage.
We self-consistently applied the same detection criteria to both the simulated satellites and the real systems when deriving our luminosity function.

\section{Observational Selection Function}\label{sec:ssf}

The observational selection function defines the detectability of a satellite as a function of its physical properties and location on the sky.
Following the convention of \citet{Koposov:2008} and \citet{Walsh:2009}, we present our observational selection function in terms of the physical parameters of heliocentric distance ($D$), absolute magnitude ($M_V$), and azimuthally averaged projected physical half-light radius ($r_{1/2}$).
The detectability of a satellite with a given set of parameters can be predicted directly from our simulations through a nearest-neighbors search \citep[e.g.,][]{1973acp..book.....K}.
However, nearest-neighbor searches can be unwieldy, and we offer two alternative parameterizations of satellite detectability based on an analytic approximation and a machine-learning classifier.
When training the machine-learning classifier, we include the local stellar density as an additional feature for predicting satellite detectability.

\subsection{Analytic Approximation}
\begin{deluxetable}{c c c c c c c}
\tabletypesize{\footnotesize}
\tablecaption{\label{tab:p50_params} Parameterization of 50\% Detectability Contour}
\tablehead{& & \colhead{DES} & & & \colhead{PS1} & \\
Distance & $A_0$ & $M_{V,0}$ & $\log_{10}(r_{1/2,0})$ & $A_0$ & $M_{V,0}$ & $\log_{10}(r_{1/2,0})$ \\
(kpc) & & (mag) & ($\log_{10}(\pc)$) &  & (mag) & ($\log_{10}(\pc)$) }
\startdata
11.3 &  21.5 & 7.8  & 3.8 & 22.8 & 7.1  & 4.0 \\ 
22.6 &  24.1 & 8.3  & 4.2 & 19.0 & 5.0  & 4.1 \\ 
45.2 &  17.2 & 5.2  & 4.3 & 14.1 & 1.8  & 4.2 \\ 
90.5 &  8.6  & 1.2  & 4.1 & 11.0 & -0.3 & 4.3 \\ 
181  &  6.6  & -1.1 & 4.1 & 7.5  & -2.2 & 4.2 \\ 
362  &  6.3  & -2.3 & 4.3 & 6.8  & -4.0 & 4.4 \\ 
\enddata
\end{deluxetable}

We first present a simple analytic approximation for the contour defining the parameters of satellites with 50\% detection probability, $\Pdet(D,M_V,r_{1/2}) = 0.5$.
We find that at fixed distance, this $P_{\rm det, 50}$ contour can be well described by
\begin{equation}
\log_{10}(r_{1/2}) = \frac{A_0(D)}{(M_V - M_{V,0}(D))} + \log_{10}(r_{1/2,0}(D)),
\end{equation}
where $r_{1/2}$ is in units of \pc, $M_V$ is in units of mag, and $D$ is in units of kpc.
This equation contains three distance-dependent constants ($A_0$, $M_{V,0}$, $r_{1/2,0}$), which were fit to the $P_{\rm det,50}$ contour in each of our six slices of distance (\tabref{p50_params}). 
In particular, $M_{V,0}$ and $\log_{10}(r_{1/2,0})$ represent asymptotic limits in absolute magnitude and physical half-light radius as a function of satellite distance.
The scale parameter, $A_0(D)$, describes the ``radius of curvature'' of the $P_{\rm det,50}$ contour at a given distance.
These fits to $P_{\rm det,50}$ are overplotted as dashed black lines on \figrefs{des_slices}{ps1_slices}.
The parameters describing $P_{\rm det,50}$ vary smoothly as a function of distance, and interpolating between them can provide a reasonable approximation for $P_{\rm det,50}$ for any distance within the range studied.

\citet{Koposov:2008} provided a similar description of satellite detectability in SDSS DR5, in terms of limiting surface brightness, $\mu_{\rm lim}$, and limiting absolute magnitude, $M_{V,{\rm lim}}$ (red dashed lines in \figrefs{des_slices}{ps1_slices}).
\citeauthor{Koposov:2008} show that this simple model is sufficient to describe the sensitivity of their search, as derived from a small sample of 8,000 simulated galaxies.
However, the functional form of this model is insufficient to fully capture the shape of the detectability contours of our search, which was derived from a much larger set of simulated galaxies.
Fitting the model of \citeauthor{Koposov:2008} to our simulations systematically underestimates our sensitivity to faint, compact satellites.
The difference between our model and that of \citeauthor{Koposov:2008} is not unexpected, since the exact shape of the detectability contour depends on the data set and search algorithm.

\subsection{Machine-learning Model}

\citet{Walsh:2009} emphasized that $P_{\rm det,50}$ serves as a useful approximation for detectability, but that it does not fully capture the intermediate region between 100\% detectability and 0\% detectability.
\CHECK{Most of the known satellites lie in the region of parameter space of intermediate detection probability, and accordingly, accurate treatment of the detection efficiency gradient in this region is an important component of the interpretation.
It is expected that many of the Milky Way satellites lie just beyond the current detection threshold \citep[\eg,][]{2014ApJ...795L..13H,Newton:2018,Jethwa:2018}, and thus any sensitivity beyond the $P_{\rm det,50}$ envelope provides valuable information on this population of faint, distant, and low-surface-brightness satellites.}

To more fully encapsulate the results of our simulations, we trained a gradient-boosted decision tree classifier to predict the detectability of a satellite based on its physical properties.
This represents a binary classification problem, where we seek to predict the relationship between a set of input features, $\vec{X}$, and a set of labels, $\vec{Y}$.
We treated each simulated satellite as a training instance, $i$, with a feature vector, $\vec{X}_i$, composed of the physical properties of the satellites. 
We labeled satellites as ``detected" ($Y_i=1$) if they satisfied the detection criteria described in \secref{detect} and ``undetected" ($Y_i=0$) otherwise.
The output of the machine-learning classifier is the probability that a satellite will be detected.\footnote{We could similarly define a regression task, which would predict the detection significance directly, but we find that the output of a classification task is more easily interpreted when deriving population statistics.}
For each survey, we classified satellites as detected/undetected, depending on whether they pass the detection criteria defined in \secref{detect}. 

We trained a gradient-boosted decision tree classifier using \code{XGBoost} \citep{Chen:2016} and \code{scikit-learn} \citep{scikit-learn} as follows:
\begin{enumerate}[wide, labelwidth=!, labelindent=0pt, itemsep=0pt]
\item Randomly split the simulated satellites into training and test sets that contain $90\%$ and $10\%$ of the simulated satellites, respectively.
\item Randomly split the training set from the previous step into $k$ hold-out cross-validation subsets. We chose $k=3$ for this analysis by performing a manual grid search over different numbers of cross-validation folds. 
\item Train a \code{XGBClassifier} using \code{GridSearchCV} to select hyperparameters that yield the best test-set classification score. Hyperparameters include the learning rate, number of trees, and maximum tree depth (see \appref{classifier}). 
\end{enumerate}
Our feature vector consisted of the absolute magnitude, the heliocentric distance, the azimuthally averaged projected half-light radius, and the stellar density at the location of each simulated satellite in the training set.

\begin{figure*}
\includegraphics[width=0.49\textwidth]{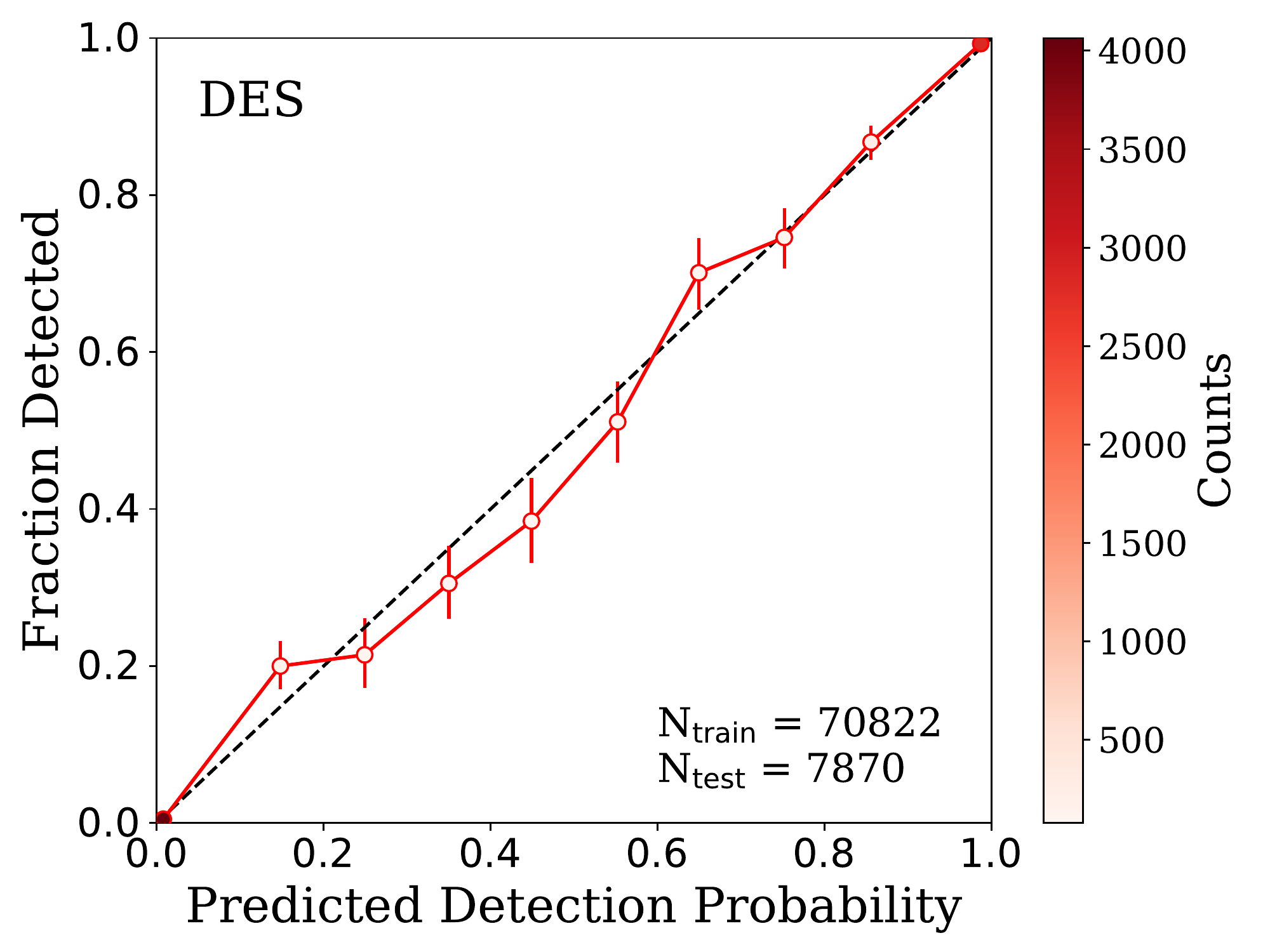}
\includegraphics[width=0.49\textwidth]{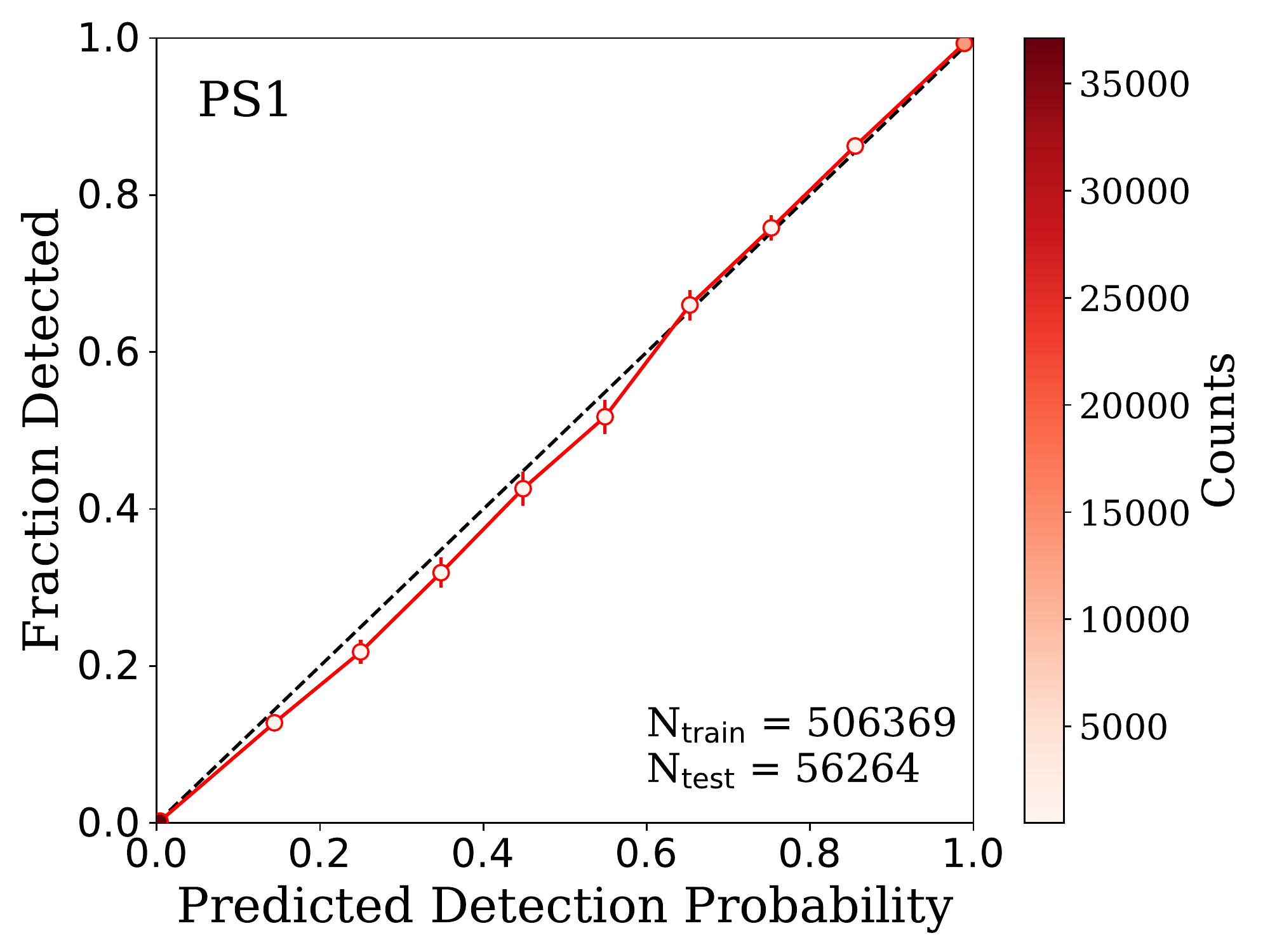}
\caption{Fraction of simulated satellites in the test set that pass our detection criteria for DES (left) and PS1 (right) versus the detection probability predicted by our ML classifier. The dashed black one-to-one line indicates perfect performance, and the colorbar indicates the number of simulated satellites in each bin of predicted detection probability. The number of simulated satellites in the training and test sets for each survey is indicated above each plot.}
\label{fig:test}
\end{figure*}

We evaluate the performance of our trained classifier using several metrics.
To assess the robustness and accuracy of the model, we evaluate the fraction of detected and undetected objects in the test set that are classified correctly.
We find that the DES classifier is \CHECK{97\%} accurate for both classes, while the PS1 classifier is \CHECK{94\%} accurate for detected objects and \CHECK{97\%} accurate for undetected objects. 
The fact that our test-set classification is accurate indicates that the training and test sets (unsurprisingly) represent the same underlying distribution. 
Moreover, because both detected and undetected objects are classified accurately, we conclude that the algorithm is not systematically biased toward either class. 
\figref{test} illustrates the true fraction of detected objects in the test set versus the detection probability predicted by the classifier. 
Even though the majority of objects are either always detected or never detected for both DES and PS1, our algorithm accurately predicts the detection probability of satellites in the intermediate regime.
The region of intermediate detection probability can be attributed to stochasticity in the distribution of stellar fluxes and spatial positions in statistical realizations of a given satellite, as well as local variations in the field population and survey characteristics.

We also trained random forest (RF) classifiers on the same sample of simulated satellites, since RFs provide easily interpretable estimates of feature importance.
We trained one RF using a minimal feature vector (absolute magnitude, heliocentric distance, and physical size) and another using a larger set of simulated satellite properties (absolute magnitude, heliocentric distance, physical size, surface brightness, \Nstars, ellipticity, sky position, and local stellar density).
We found that the RF model trained on the full feature vector was slightly more accurate compared to the three-feature RF, although both were biased for high- and low-detection probability objects with respect to our nominal algorithm. 
The relatively small improvement from using the full feature vector gives us confidence that our nominal feature vector adequately captures much of the necessary information to predict detectability.
We note that retraining our selection-function classifier using additional features is straightforward. 

We publicly distribute the trained classifiers to encapsulate the sensitivity of DES and PS1.\footnote{\url{https://github.com/des-science/mw-sats}}
These classifiers can be used to predict the probability that a satellite would be detected as a function of its physical parameters and location on the sky.
These detection probabilities can be used to predict the number of observed galaxies from a model of the full underlying Milky Way satellite galaxy population (e.g., Paper II).

\section{Satellite Luminosity Function}\label{sec:lf}

The observational selection functions derived for DES and PS1 allow us to predict the detectability of a satellite, given its physical properties. 
As a simple demonstration, we apply these selection functions to estimate the luminosity function of Milky Way satellite galaxies assuming that satellites are distributed isotropically and that the physical properties of the observed satellites are representative of the total population. 
The results of this analysis may be compared to similar analyses based on data from SDSS \citep[e.g.,][]{Koposov:2008,2008ApJ...688..277T,Walsh:2009}; however, recent observations suggest that a more complex modeling framework is necessary \citep[e.g.][]{Jethwa:2018,Newton:2018,Nadler:2019a}.
In Paper II, we perform a more rigorous analysis of the Milky Way satellite population based on cosmological simulations, which includes the effects of satellite disruption, galaxy formation efficiency, and the presence of the LMC.

Following a procedure similar to that of \citet{Koposov:2008}, we used the trained machine-learning classifier to derive the probability that each satellite will be detected as a function of Galactocentric position.
For each confirmed and candidate dwarf galaxy (indexed by $i$) in the DES and PS1 footprints (class $\geq 3$ in \tabref{all_mw_sats}), we calculated the probability that a galaxy with the same parameters would be found within the survey volume,
\begin{equation}
C_{V,i} = \frac{\displaystyle \int\int_{R_{\rm min}}^{R_{\rm max}} \Pdet(D(\vec{r}), \rho_\star(\vec{r}), M_{V,i}, r_{1/2,i})\,n(r)\,r^2 \dd{r}\,\dd{\Omega}}{\displaystyle \int \int_{R_{\rm min}}^{R_{\rm max}} n(r)\,r^2 \dd{r}\,\dd{\Omega}},\label{eqn:cvi}
\end{equation}
where \Pdet is the probability that a satellite with absolute magnitude $M_{V,i}$ and physical half-light radius $r_{1/2,i}$ will be detected at a heliocentric distance $D$ and local stellar density $\rho_\star$, which are functions of the Galactocentric location of the satellite, $\vec{r}$.\footnote{We assume the Galactocentric distance of the Sun is $8.3\kpc$ \citep{Gillessen:2009}.}
Following \citet{Koposov:2008}, we assumed that the Galactocentric radial distribution of satellites follows a cored NFW profile, $n(r) \propto  r^{-2} (r + r_c)^{-1}$, where a core radius of $r_c = 20\kpc$ was adopted to prevent divergence in the inner regions. 
Moreover, we adopted an inner cutoff of $R_{\rm min} = 20\kpc$ in our radial integration to crudely estimate the tidal effects of the Galactic disk, which generally disrupts subhalos that pass within this radius \citep[\eg,][]{Garrison-Kimmel:2017}.
We integrate the satellite population out to $R_{\rm max} = 300 \kpc$, which is consistent with similar analyses in the literature \citep[\eg,][]{2014ApJ...795L..13H,Jethwa:2018,Newton:2018,Nadler:2019a} and is roughly comparable to the virial radius of the Milky Way \citep[e.g.,][]{Garrison-Kimmel:2014}.
The angular distribution of satellites is assumed to be isotropic in Galactocentric coordinates and is transformed into celestial equatorial coordinates to estimate the local stellar density.

To incorporate information from both surveys, we defined a weight for each satellite,
\begin{equation}
w_i = \frac{1}{C_{V,i,{\rm DES}} + C_{V,i,{\rm PS1}}},
\end{equation}
which combines the volumetric correction factors for the two surveys.
To avoid double counting, we removed the area from PS1 that overlapped with DES.

\begin{figure*}[t]
\centering
\includegraphics[width=0.49\textwidth]{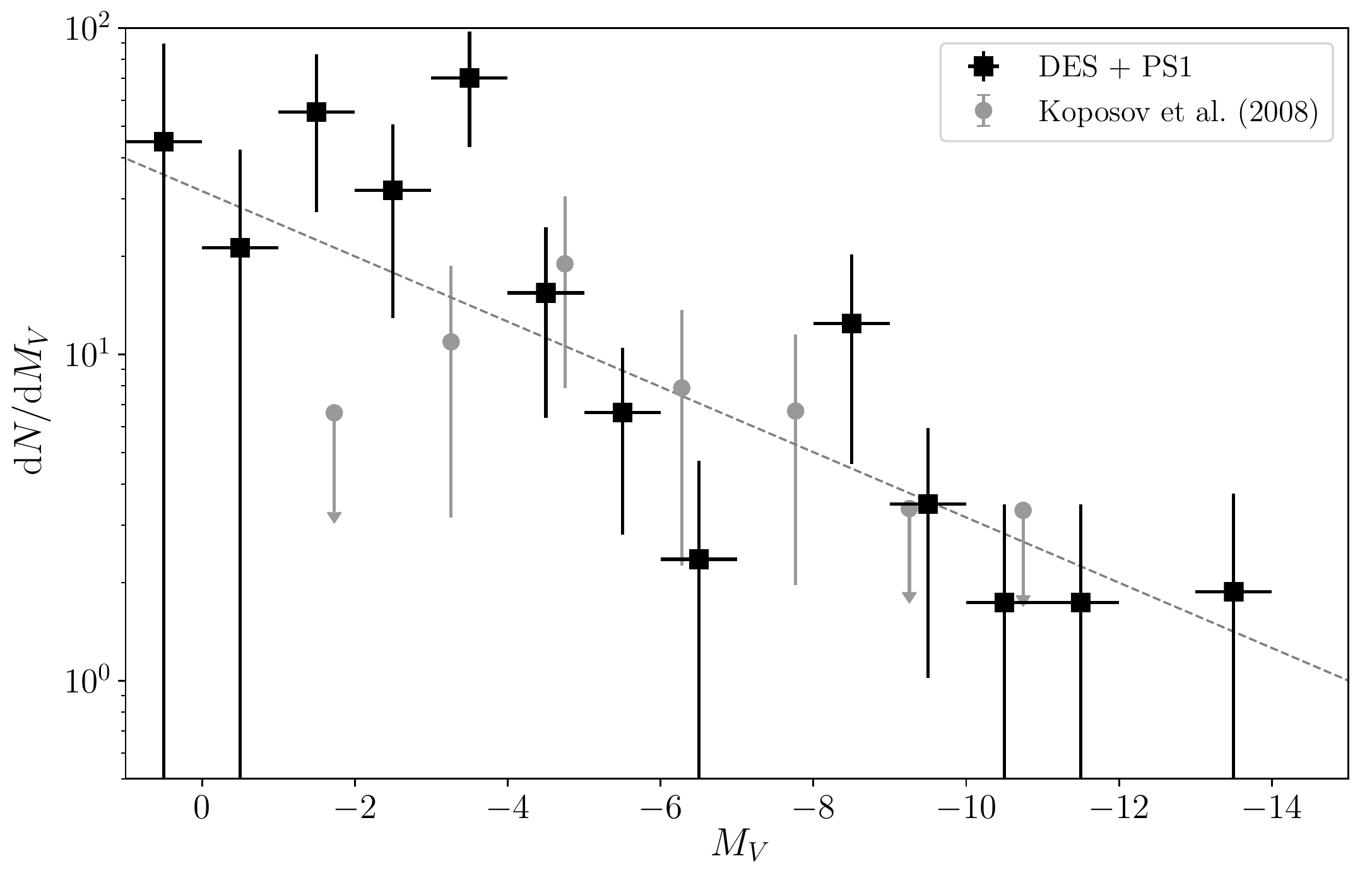}
\includegraphics[width=0.49\textwidth]{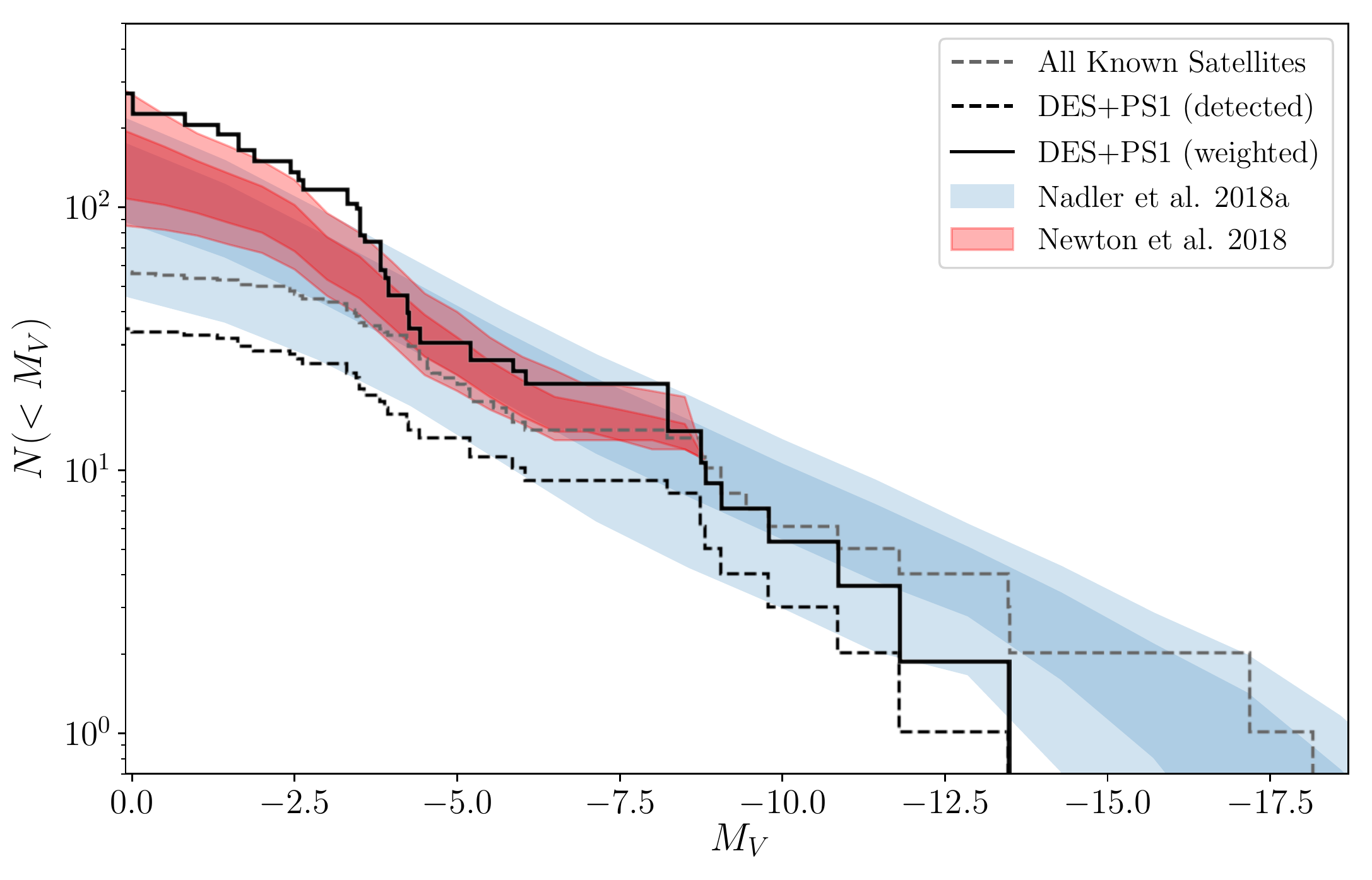}
\caption{\label{fig:dNdMv} 
Luminosity functions of Milky Way satellite galaxies within $R_{\rm vir} = 300 \kpc$ inferred from our analysis under the assumptions of an isotropic satellite distribution and a cored NFW radial distribution. 
{\it (Left)}: Differential number of satellites within bins of absolute V-band magnitude.
This calculation uses satellites detected with our selection threshold of $\sqrt{\TS} > 6$ and $\SIG > 6$.
The observed satellites have been volume corrected using the observational selection functions for DES and PS1 derived here.
Uncertainties are estimated from the Poisson error on the number of satellites in each bin.
The power law function $\mathrm{d}N / \mathrm{d}M_V = 10\times10^{0.1(M_V+5)}$ from \citet{Koposov:2008} is shown as a dashed gray line.
{\it (Right)}: Cumulative number of Milky Way satellites. All known satellites are shown by the dashed gray line, while the satellites detected in our search of DES Y3A2 and PS1 DR1 are shown by the black dashed line.
The solid black line shows detected satellites in DES and PS1 weighted by the observable volume correction for each satellite.
Also shown are results from \citet{Newton:2018} and \citet{Nadler:2019a} that combine the observed satellite population with numerical simulations of dark matter and galaxy formation. 
}
\end{figure*}

The resulting volume-weighted Milky Way satellite luminosity function is shown in \figref{dNdMv}. 
The left panel of \figref{dNdMv} shows the differential volume-weighted number of satellites in bins of absolute magnitude, $\mathrm{d}N/\mathrm{d}M_V$.
Both PS1 and DES are largely complete out to the virial radius for satellites with \CHECK{$M_V \lesssim -7.5$ and $r_{1/2} \lesssim 200\pc$}. 
Thus, for brighter satellites, the volume correction amounts to an area correction due to incomplete sky coverage.
Uncertainties on the weighted number of satellites in each bin are calculated as $\sqrt{ \sum_i w_i^2}$, which is equivalent to a weighted Poisson uncertainty.

For comparison, we also plot the differential luminosity function measured by \citet{Koposov:2008} using data from SDSS DR5 (gray points with uncertainties).
When compared to previous analyses of SDSS \citep{Koposov:2008,Walsh:2009}, our analysis covers $\roughly 3$ times the sky area and includes $\roughly 3$ times as many satellite galaxies.
This allows us to extend the direct calculation of the differential Milky Way satellite luminosity function to both brighter and fainter systems.
To guide the eye, we include the power-law model, $\mathrm{d}N/\mathrm{d}M_V = 10 \times 10^{0.1 ( M_V + 5)}$, described by \citet{Koposov:2008}.

The right panel of \figref{dNdMv} shows the cumulative number of satellites brighter than a given absolute magnitude.
We show both the total number of known satellites (gray dashed line) and the number of satellites detected in our search of DES and PS1 (black dashed line).
We correct the detected satellite curve by the volumetric weights described previously to yield the volume-corrected cumulative satellite luminosity function (black solid line).
This estimate of the satellite luminosity function assumes that satellites are distributed isotropically, that the radial distribution of satellites goes as $r^{-3}$ at large radii, and that the physical sizes of the known satellites are representative of the population as a whole.
This analysis  predicts that the Milky Way contains \CHECK{$\roughly 270$} satellite galaxies within 300\kpc with $M_V \lesssim 0$.
The resulting cumulative number of satellites is slightly higher than predictions from some cosmological simulations \citep[e.g.,][]{Newton:2018,Nadler:2019a}, but lower than others \citep[e.g.,][]{Kelley:2018}.
The sharp upturn in $N(< M_V)$ around $M_V \sim -4$ comes from recent discoveries in the Southern Hemisphere combined with increasing volumetric weights at these faint magnitudes.
This upturn at faint magnitudes has been attributed to the existence of a population of faint satellites associated with the LMC \citep{Jethwa:2018,Newton:2018} and/or reionization physics \citep{Bose:2018}.
We discuss the importance of modeling the LMC satellite system in more detail in Paper II.
 
The analysis described here has few \emph{explicit} modeling choices (\ie, the functional form of $n(r)$); however, it contains several \emph{implicit} assumptions about the satellite population.
First, we have assumed that the physical properties of the undetected satellite population are consistent with the population of detected satellites.
Such an assumption breaks down if a large population of extremely low-surface-brightness systems exists (\eg, satellites like Crater II and Antlia II).
A population of large, luminous, and distant Milky Way satellites has been suggested to resolve the observed radial distribution discrepancy between the satellite populations of the Milky Way and Andromeda \citep[e.g.,][]{Yniguez:2014,Samuel:2020}.
In addition, we assumed that the intrinsic properties of satellites (\eg, luminosity and half-light radius) are independent of heliocentric distance.
Again, this assumption may break down due to the larger influence of baryonic physics on the properties of satellites close to the Milky Way \citep[e.g.,][]{Nadler:2018}.
Finally, we made the simplifying assumption that the satellite population is isotropic, while observations suggest that this is very unlikely to be true for the Milky Way \citep[e.g.,][]{Pawlowski:2012,Koposov:2015,Bechtol:2015,Drlica-Wagner:2015} and Andromeda \citep[e.g.,][]{2013Natur.493...62I}.
These assumptions strongly motivate the more rigorous analysis in Paper II, which simultaneously includes the effects of satellite disruption (enhanced by the presence of the Galactic disk), uncertainties in the efficiency of galaxy formation in low-mass halos, and the influence of the LMC.

\section{Conclusions}\label{sec:conclusions}

We have conducted a search for low-luminosity Milky Way satellite galaxies over $\roughly 24,000 \deg^2$ ($\roughly 75\%$ of the high-Galactic-latitude sky) using data from DES Y3A2 and PS1 DR1.
We recover most of the satellites previously discovered in the DES and PS1 footprints, though a number of low-luminosity satellites are beyond the sensitivity limit of our search.
We set a detection threshold intended to minimize the number of false-positive candidates, and no new, high-confidence satellite galaxy candidates were discovered.
The only significant new candidate is a compact, low-luminosity outer-halo star cluster (\appref{candidates}).

We determined the sensitivity of our search by simulating the resolved stellar populations of Milky Way satellite galaxies, injecting them into the DES and PS1 catalogs, and running \emph{the same} satellite detection algorithms that were applied to the real data.
We quantified the observational sensitivity of our search in terms of satellite properties (\ie, absolute magnitude, physical size, distance, and local stellar density) and provide both analytic and machine-learning models of the observational selection function.
Finally, we demonstrated the application of our observational selection function to derive a data-driven luminosity function from the observed Milky Way satellite galaxy population.
By encapsulating the observational selection function in a flexible and accurate machine-learning model, we facilitate more rigorous statistical-inference-based approaches to extract the properties of galaxy formation and dark matter physics.

Deep, multi-band optical imaging over the entire sky at depths fainter than $g \sim 23$ is now within reach.
DES covers only one-sixth of the sky accessible to DECam, and the past several years have seen an active community campaign to complete contiguous DECam coverage of the entire southern sky.
Programs like MagLiteS \citep{Drlica-Wagner:2016hwk, Torrealba:2018a}, BLISS \citep{Mau:2019}, DECaLS \citep{Dey:2019}, and DELVE \citep{Mau:2019b}\footnote{\url{https://delve-survey.github.io}} are actively collecting, processing, and mining these data for fainter and more distant satellites.
Meanwhile,  HSC SSP on the 8.2-m Subaru telescope will achieve $r \sim 26$ over $\roughly 1,400 \deg^2$, thereby extending the search for the faintest galaxies to unprecedented distances \citep{Homma:2016,Homma:2017,Homma:2019}.
In the early 2020s, the Legacy Survey of Space and Time (LSST) conducted from the Vera C.\ Rubin Observatory will expand this depth of coverage to the entire southern sky.
The power of these upcoming surveys, combined with advanced modeling techniques, promise to shed new light on the darkest galaxies.

\section{Acknowledgments}\label{sec:ack}
This research was supported in part by the National Science Foundation (NSF) under Grant No.\ NSF PHY-1748958 through the Kavli Institute for Theoretical Physics program ``The Small-Scale Structure of Cold(?)\ Dark Matter,'' and grant no.\ NSF DGE-1656518 through the NSF Graduate Research Fellowship received by E.O.N. 
S.M. is supported by the University of Chicago Provost's Scholar Award.

This research made use of computational resources at SLAC National Accelerator Laboratory, a U.S.\ Department of Energy Office; the authors thank the support of the SLAC computational team.
This research has made use of the WEBDA database, operated at the Department of Theoretical Physics and Astrophysics of the Masaryk University.

Funding for the DES Projects has been provided by the U.S. Department of Energy, the U.S. National Science Foundation, the Ministry of Science and Education of Spain, 
the Science and Technology Facilities Council of the United Kingdom, the Higher Education Funding Council for England, the National Center for Supercomputing 
Applications at the University of Illinois at Urbana-Champaign, the Kavli Institute of Cosmological Physics at the University of Chicago, 
the Center for Cosmology and Astro-Particle Physics at the Ohio State University,
the Mitchell Institute for Fundamental Physics and Astronomy at Texas A\&M University, Financiadora de Estudos e Projetos, 
Funda{\c c}{\~a}o Carlos Chagas Filho de Amparo {\`a} Pesquisa do Estado do Rio de Janeiro, Conselho Nacional de Desenvolvimento Cient{\'i}fico e Tecnol{\'o}gico and 
the Minist{\'e}rio da Ci{\^e}ncia, Tecnologia e Inova{\c c}{\~a}o, the Deutsche Forschungsgemeinschaft and the Collaborating Institutions in the Dark Energy Survey. 

The Collaborating Institutions are Argonne National Laboratory, the University of California at Santa Cruz, the University of Cambridge, Centro de Investigaciones Energ{\'e}ticas, 
Medioambientales y Tecnol{\'o}gicas-Madrid, the University of Chicago, University College London, the DES-Brazil Consortium, the University of Edinburgh, 
the Eidgen{\"o}ssische Technische Hochschule (ETH) Z{\"u}rich, 
Fermi National Accelerator Laboratory, the University of Illinois at Urbana-Champaign, the Institut de Ci{\`e}ncies de l'Espai (IEEC/CSIC), 
the Institut de F{\'i}sica d'Altes Energies, Lawrence Berkeley National Laboratory, the Ludwig-Maximilians Universit{\"a}t M{\"u}nchen and the associated Excellence Cluster Universe, 
the University of Michigan, the NSF’s National Optical-Infrared Astronomy Laboratory, the University of Nottingham, The Ohio State University, the University of Pennsylvania, the University of Portsmouth, 
SLAC National Accelerator Laboratory, Stanford University, the University of Sussex, Texas A\&M University, and the OzDES Membership Consortium.

Based in part on observations at Cerro Tololo Inter-American Observatory, NSF’s National Optical-Infrared Astronomy Laboratory, which is operated by the Association of 
Universities for Research in Astronomy (AURA) under a cooperative agreement with the National Science Foundation.

The DES data management system is supported by the National Science Foundation under Grant Numbers AST-1138766 and AST-1536171.
The DES participants from Spanish institutions are partially supported by MINECO under grants AYA2015-71825, ESP2015-66861, FPA2015-68048, SEV-2016-0588, SEV-2016-0597, and MDM-2015-0509, 
some of which include ERDF funds from the European Union. IFAE is partially funded by the CERCA program of the Generalitat de Catalunya.
Research leading to these results has received funding from the European Research
Council under the European Union's Seventh Framework Program (FP7/2007-2013) including ERC grant agreements 240672, 291329, and 306478.
We  acknowledge support from the Brazilian Instituto Nacional de Ci\^encia
e Tecnologia (INCT) e-Universe (CNPq grant 465376/2014-2).

This manuscript has been authored by Fermi Research Alliance, LLC under Contract No. DE-AC02-07CH11359 with the U.S. Department of Energy, Office of Science, Office of High Energy Physics. The United States Government retains and the publisher, by accepting the article for publication, acknowledges that the United States Government retains a non-exclusive, paid-up, irrevocable, world-wide license to publish or reproduce the published form of this manuscript, or allow others to do so, for United States Government purposes.

\bibliographystyle{apj}
\bibliography{main}

\appendix

\section{Data Selection}
\label{app:ps1}
\label{app:des}

\subsection{DES Y3A2}

The DES data were selected from the Y3A2 internal release of the GOLD catalog (v2.0) accessed via a bulk download from the \code{easyaccess} SQL command line interpreter \citep{CarrascoKind:2019}. 
We removed objects with \code{FLAGS\_GOLD} \& 0b111100, which identifies SOF fit failures, objects with SExtractor $\code{FLAGS} > 3$, objects with bad pixels in their isophotes \code{IMAFLAGS\_ISO !=} 0, objects characterized as bright blue color outliers \citep[Section 6.2 of][]{Drlica-Wagner:2018}, and other extreme color outliers \citep[Section 6.2 of][]{Drlica-Wagner:2018}.
Magnitudes were corrected for a CCD-dependent magnitude adjustment to the zero-points (\code{DELTA\_MAG\_V3}), an SED-dependent chromatic correction (\code{DELTA\_MAG\_CHROM}), and extinction using the extinction maps from \citet{Schlegel:1998} (\code{A\_SED\_SFD98}).
Star--galaxy classification used the \code{EXTENDED\_CLASS\_MASH\_SOF} variable, with stars selected with $0 \leq \code{EXTENDED\_CLASS\_MASH\_SOF} \leq 2$. 
The construction of the extended classification variable is described in Section 2 of \citep{Shipp:2018}.
Briefly, this classifier uses the SOF size parameter \code{CM\_T} (and associated error) to classify most objects.
When the SOF parameters are unavailable (due to a small fraction of objects where the SOF fit fails), the weighted average of the single-epoch measurements \code{WAVG\_SPREAD\_MODEL} (and associated error) are used for classification.
For faint objects where both SOF and WAVG values are unavailable, classification is performed with the \code{SPREAD\_MODEL} (and associated uncertainty) parameters derived from the coadded images.

\subsection{PS1 DR1}

The PS1 DR1 data were downloaded from the MAST CasJobs server using queries similar to the following example.
Note that the query selects objects in an interval in \code{objid}. 
The full bulk download is $\roughly10{,}000$ slices in \code{objid} designed to have approximately equal numbers of objects per slice. 
Each slice is effectively a narrow interval in declination. 
In a post-processing step, the objects were partitioned into a \healpix grid for rapid access during the search phase.

\begin{verbatim}
  SELECT
    f.raMean, f.decMean, 
    f.objID, f.uniquePspsOBid,
    f.objInfoFlag, f.qualityFlag,
    f.nStackDetections, f.nDetections, f.ng, f.nr, f.ni,
    f.gFPSFFlux, f.rFPSFFlux, f.iFPSFFlux, 
    f.gFPSFFluxErr, f.rFPSFFluxErr, f.iFPSFFluxErr,
    f.gFKronFlux, f.rFKronFlux, f.iFKronFlux,
    f.gFKronFluxErr, f.rFKronFluxErr, f.iFKronFluxErr,
    f.gFlags, f.rFlags, f.iFlags,
    s.gInfoFlag, s.rInfoFlag, s.iInfoFlag,
    s.gInfoFlag2, s.rInfoFlag2, s.iInfoFlag2,
    s.gInfoFlag3, s.rInfoFlag3, s.iInfoFlag3,
    s.primaryDetection, s.bestDetection
  INTO MyDB.ps1_dr1_objid_bin_09991_10000_v02
  FROM ForcedMeanObjectView f
  JOIN StackObjectThin s on f.objid = s.objid
  WHERE f.objid BETWEEN 212390000000001521 AND 212590000000007936
  AND f.qualityFlag & 16 > 0;
\end{verbatim}

We remove duplicate objects by requiring

\begin{verbatim}
    sel = (data["primaryDetection"] == 1)
\end{verbatim}

\noindent and apply additional selection criteria to filter the high-quality catalog objects.

\begin{verbatim}
    sel &= (data["nStackDetections"] > 1)
    sel &= (data["nDetections"] > 0)
    sel &= (data["gInfoFlag"] >= 0)
    sel &= (data["rInfoFlag"] >= 0)
    sel &= (data["iInfoFlag"] >= 0)
    sel &= ((data["gInfoFlag"] & (8 + 2048)) == 0)
    sel &= ((data["rInfoFlag"] & (8 + 2048)) == 0)
    sel &= ((data["iInfoFlag"] & (8 + 2048)) == 0)
    sel &= ((data["gInfoFlag2"] & (8192 + 4194304)) == 0)
    sel &= ((data["rInfoFlag2"] & (8192 + 4194304)) == 0)
    sel &= ((data["iInfoFlag2"] & (8192 + 4194304)) == 0)
\end{verbatim}

\noindent These criteria were determined by comparing object detections in PS1 to detections in deeper imaging data from DES and HSC, to identify the quantities that correlate with higher rates of unmatched PS1 objects.

Our star--galaxy selection was performed using the measured aperture and PSF magnitudes in $i$-band:
\begin{verbatim}
    sel  = ((data["iFPSFMag"] - data["IFKronMag"]) < 0.05)
    sel |=  (data["iFPSFMag"] == -999.)
    sel |= ((data["iFPSFMag"] - data["IFKronMag"]) > 4.0)
\end{verbatim}
The $i$-band was chosen due to its superior PSF and depth.

For the \simple analysis, we also require moderate signal-to-noise object detections ($\mathrm{S/N} > 10$) to ensure a sample of high-confidence stars:
\begin{verbatim}
    sel |= ((data["rFPSFMagErr"] < 0.1)
\end{verbatim}
While this S/N selection limited the number of faint stars used in the search (\figref{photo_err}), it significantly reduced the number of spurious candidates returned by the \simple algorithm. The S/N selection was not applied to the \ugali search on PS1.

\section{Validation of the Parameter Space for Simulated Satellites}
\label{app:mu_ng22}

The ability to detect satellites depends largely on the surface brightness and number of resolved member stars recovered by a survey \citep[\eg,][]{Walsh:2009}.
These observed properties depend both on the physical characteristics of the satellite (\eg, luminosity, size, distance) and the characteristics of the survey (\eg, completeness, coverage).
For DES and PS1, we parameterize these two attributes using the average surface brightness within the half-light radius, $\mu$, and the number of satellite member stars brighter than $g = 22$, $N(g < 22$).
The \Nstars parameter is challenging to compare directly to models of galaxy formation due to its dependence on the survey completeness, but is readily obtained from the numerical simulations of satellites, and is an effective means to gauge whether a given satellite could be detected.

\begin{figure*}[t]
\centering
\includegraphics[width=\textwidth]{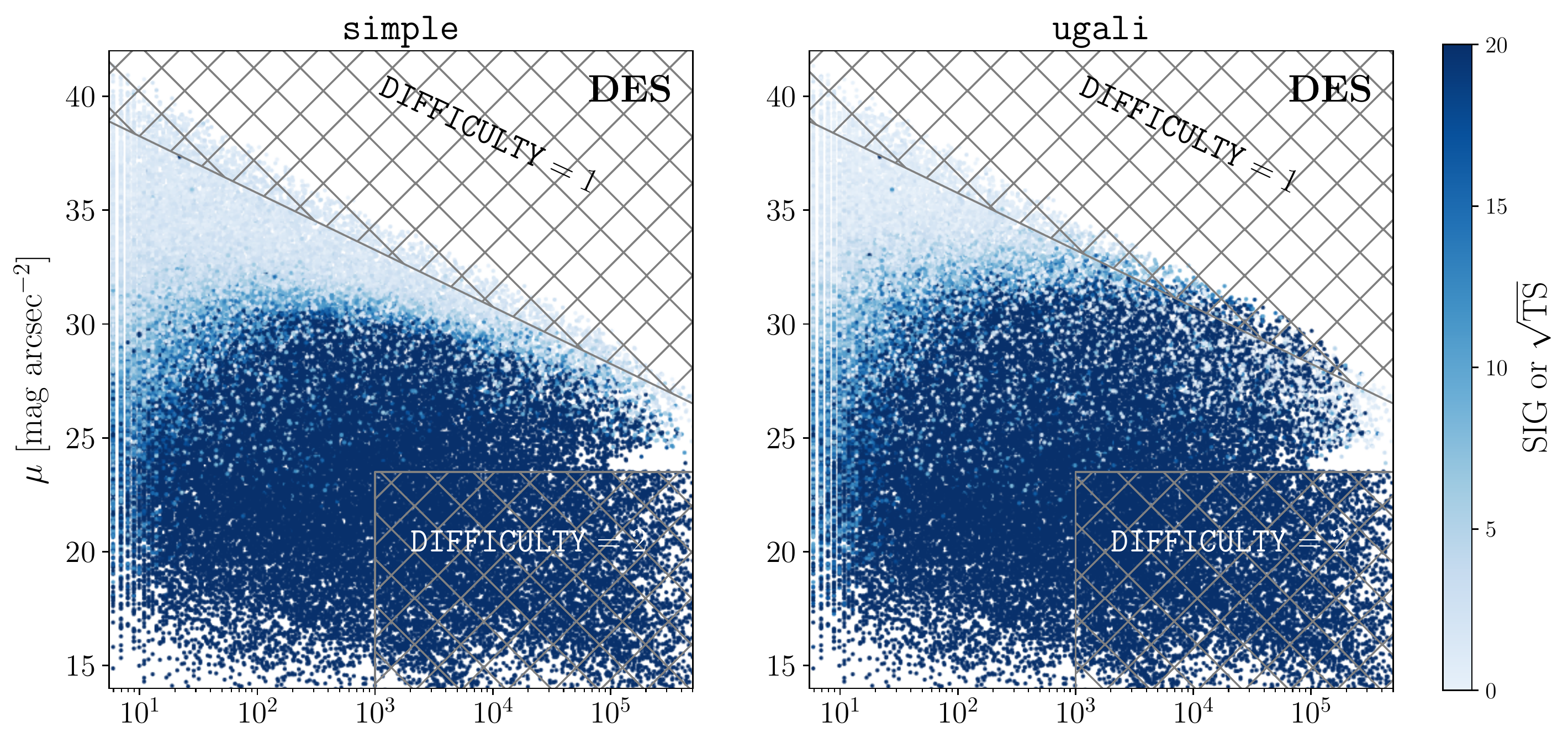}\\
\includegraphics[width=\textwidth]{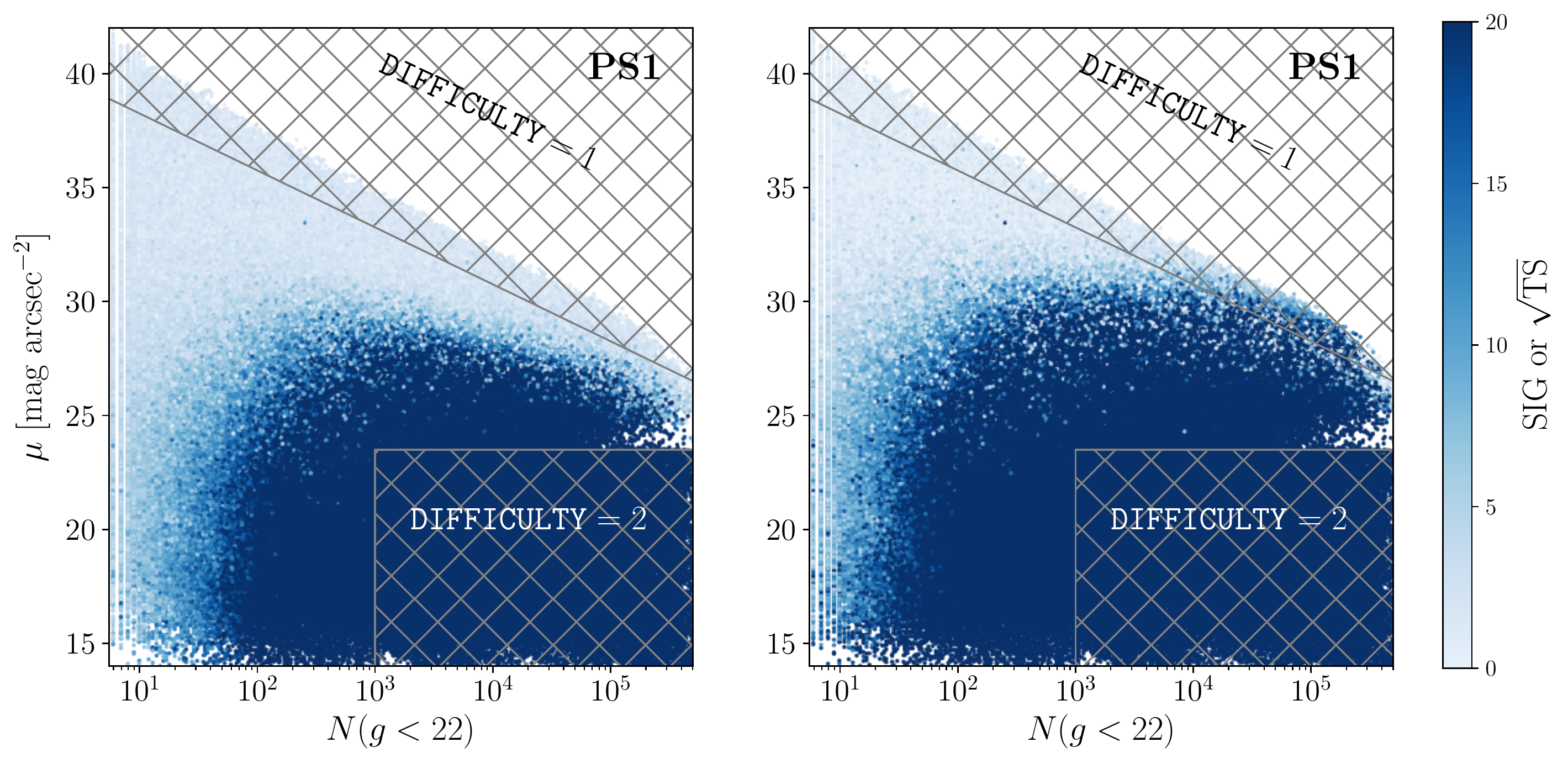}
\caption{Detection significance of simulated satellites as a function of surface brightness, $\mu$, and the number of member stars brighter than $g = 22$, $N(g < 22$). 
The top (bottom) row shows results for DES Y3A2 (PS1 DR1) with each point corresponding to a simulated satellite. 
The colorbar indicates the detection significance corresponding to the \simple \SIG value (left column) or the \ugali $\sqrt{\TS}$ (right column). 
Satellites with $a_h > 5 \deg$ are flagged with \code{DIFFICULTY = 1} and are not used in our training.
Satellites in the region labeled \code{DIFFICULTY = 2} have both high surface brightness and many resolved stars. 
We assume satellites in this region are always detected and are therefore not simulated for computational efficiency.
}
\label{fig:mu_vs_N24_panels}
\end{figure*}

To assess the sensitivity of our two search algorithms (\simple and \ugali) applied to our two data sets (DES Y3A2 and PS1 DR1), we show the recovered significance of satellites as a function of their surface brightness, $\mu$, and the number of bright resolved member stars, \Nstars, in \figref{mu_vs_N24_panels}.
The detectability of satellites is easily characterized in this parameter space; in particular, for DES Y3A2 (PS1 DR1), satellites brighter than $\mu \sim 30 \magn \asec^{-2}$ ($\mu \sim 28 \magn \asec^{-2}$) with more than $\Nstars \gtrsim 10^{2}$ resolved member stars are reliably detected, while lower surface brightness systems or those with fewer resolved stars are detected less efficiently.
The cross-hatching indicates regions where satellites start to become large enough that sky projection effects come into consideration ($a_h > 5 \deg$; \code{DIFFICULTY = 1}) or have so many stars that direct simulation becomes too computationally intensive ($\Nstars > 10^3$ and $\mu < 23.5 \magn \asec^{-2}$; \code{DIFFICULTY = 2}).
We only used simulations with \code{DIFFICULTY = 0} when training the machine-learning model, though the classifier accurately predicts the detectability of simulated satellites in regions of parameter space with \code{DIFFICULTY != 0}.

\section{Maximum-likelihood Formalism}
\label{app:like}

The extended maximum-likelihood formalism \citep[\eg,][]{Orear:1958,Barlow:1990} is widely used in physics and astronomy as a means to perform statistical searches for overdensities in a Poisson random field with unknown signal normalization (\eg, galaxy cluster searches as described in \citealt{Kepner:1999} or \citealt{Rykoff:2012}).
In this appendix, we describe the application of this formalism to the search for stellar overdensities, as implemented in the \ugali software package.\footnote{\url{https://github.com/DarkEnergySurvey/ugali}}

We assume that the angular and magnitude distribution of stars in a survey can be described by a Poisson realization of the underlying stellar density field. 
Dividing the survey into bins, we can write the binned Poisson likelihood, $\mathcal{L}$, as 
\begin{equation}
    \label{eqn:like}
    \mathcal{L} = \prod_i^{\rm bins} P_i = \prod_i^{\rm bins} \frac{{n_i}^{k_i} e^{-n_i}}{k_i !}
\end{equation}
where in bin $i$, $P_i$ is the Poisson probability to observe $k_i$ stars, given a model expectation of $n_i$ stars. 
This likelihood is valid for binning over any domain, but for our specific application to satellite galaxy searches, we commonly utilize two spatial dimensions (\eg, right ascension and declination) and two magnitude dimensions (\eg, $g$ and $r$). 
For numerical simplicity, we generally work with the logarithm of the likelihood
\begin{equation}
    \label{eqn:loglike}
    \log \mathcal{L} = \sum_i^{\rm bins} \big\{ - n_i + k_i \log(n_i) - \log{(k_i!)} \big\}.
\end{equation}
The last term in this equation, $\log{(k_i!)}$, does not depend on any of the model parameters and can be safely discarded as an additive constant.

The number of model-predicted stars in a bin is the integral of the probability density function (PDF) of the model over that bin, $n_i = \int_i m(\lambda,\vect{\theta})\, \dd{\mathcal{V}}$. 
In this equation, the integral is over the observable ``volume'' of the bin, while $\lambda$ and $\vect{\theta}$ are parameters of our model.
We explicitly formulate our model for the stellar counts in terms of a contribution from the satellite and a contribution from Milky Way field stars,
\begin{equation}
    \label{eqn:model}
    m(\lambda, \vect{\theta}) = \lambda s(\vect{\theta}) + b(\vect{\theta}).
\end{equation}
In the preceding equations, we have been explicit about the dependence of $m$ on the signal normalization parameter, $\lambda$, which we call the ``richness''.\footnote{The term ``richness'' is inherited from the analysis of galaxy clusters \citep[\eg,][]{Abell:1958}.}

In the limit of infinitesimally small bins, each bin contains either zero or one star ($k_i \in \{0,1\}$), and the log-likelihood can be expressed as
\begin{equation}
  \label{eqn:loglike2}
  \log \mathcal{L} = -\sum_i^{\rm empty} n_i -\sum_i^{\rm filled} n_i + \sum_i^{\rm filled} {\log{(n_i)}}
\end{equation}
By definition, the sum over empty and filled bins is equivalent to the sum over all bins.
Thus, the first two terms can be written as the integral of the model over all observable space,
\begin{align}
  \sum_i^{\rm bins} n_i &= \int_{\mathcal{V}}  m(\lambda,\vect{\theta}) \, \dd{\mathcal{V}} \\
                     &= \lambda N_s + N_b.
\end{align}
In the last line, we have used $N_s$ and $N_b$ to represent the total number of expected satellite stars and Milky Way field stars, respectively.
The only filled bins will be located on the observed stars. 
Thus, the last term in \eqnref{loglike2} can be rewritten as a sum over stars, where we replace the model-predicted counts with the model PDF evaluated at the location of each star,
\begin{equation}
  \sum_i^{\rm filled} \log{(n_i)} = \sum_j^{\rm stars} \log(\lambda s_j + b_j).
\end{equation}
where $j$ indexes over stars, and we use $s_j,\,b_j$ as a shorthand for the components of the model PDF evaluated at the location of each star.
Substituting back into \eqnref{loglike2} yields
\begin{equation}
  \label{eqn:ulike}
  \log \mathcal{L} = - \lambda N_s - N_b + \sum_j^{\rm stars}\log(\lambda s_j + b_j),
\end{equation}
which is conventionally referred to as the unbinned Poisson log-likelihood function.

When searching for stellar overdensities, we are interested in maximizing the likelihood with respect to the richness parameter, $\lambda$. 
Differentiating \eqnref{ulike} with respect to $\lambda$ gives
\begin{equation}
  \label{eqn:dlike}
    \frac{ \dd{\log{\mathcal{L}}} }{\dd{\lambda}} =  - N_s + \sum_j^{\rm stars} \frac{s_j}{\lambda s_j + b_j}.
\end{equation}
The maximum-likelihood estimator for the richness, $\lambda = \hat{\lambda}$, will occur when the derivative of the likelihood is zero; thus,
\begin{equation}
  \label{eqn:maxlike}
  \hat{\lambda} N_s = \sum_j^{\rm stars} \frac{\hat{\lambda} s_j}{\hat{\lambda} s_j + b_j}.
\end{equation}
The left-hand side of this equation represents the number of observable stars predicted by the signal model, while the right-hand side represents the sum of the probabilities that each star belongs to the signal distribution,
\begin{equation}
    p_j \equiv \frac{\lambda s_j}{\lambda s_j + b_j}.
\end{equation}
If we care only about the dependence of the likelihood on $\lambda$, we can rewrite \eqnref{ulike} in terms of the observable fraction and the membership probabilities,\footnote{We note a typo in \citet{Bechtol:2015} that omitted the ``$\log$'' from the right-hand side of \eqnref{plike}.}
\begin{equation}
  \label{eqn:plike}
  \log{\mathcal{L}} = -\lambda N_s - \sum_j^{\rm stars} \log(1 - p_j).
\end{equation}
This formulation makes explicit the dependence of the likelihood on the membership probabilities, but discards terms that depend on the background model alone.

From \eqnref{model} it is clear that there is a degeneracy between the normalization of the signal PDF, $s$, and the richness, $\lambda$. 
Some authors \citep[\eg,][]{2008ApJ...684.1075M} choose to normalize the signal PDF to unity over the \emph{observed} space, thus requiring an observational correction to predict the number of model stars below the detection threshold of a survey.
In contrast, the \ugali framework normalizes the signal PDF to unity over the \emph{entire} observable domain, $\int_{\rm all} s(\vect{\theta})\, \dd{\mathcal{V}} = 1$ and defines the observable fraction of the signal, $f$, at a given location in a survey. 
Thus, the number of observable stars is
\begin{align}
\lambda N_s = \lambda \int_{\rm all}{ f s(\vect{\theta})\, \dd \mathcal{V} } = \lambda f.
\end{align}
This definition allows us to interpret $\lambda$ as the \emph{total} number of stars in the satellite, rather than the number of observed stars.

\section{Remaining Candidates}
\label{app:candidates}

\begin{figure*}[t]
\centering
\includegraphics[width=0.8\textwidth]{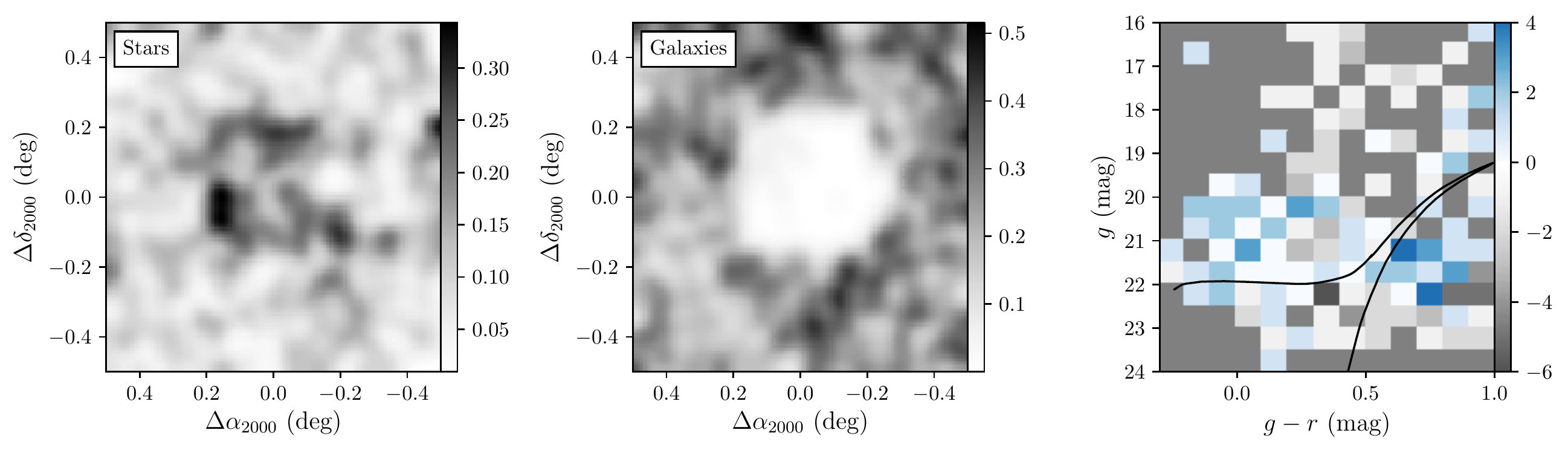}
\caption{\label{fig:skycell} 
Artifact resulting from the bright star (BD$+$18 240) biasing the processing and catalog creation for a \code{stack} image in a single PS1 DR1 \code{skycell}. Artifacts like this were identified visually in the PS1 DR1 candidate list, and regions around the affected \code{skycells} were masked.
Similar figure to \figref{cluster}, but with the stellar and galactic densities convolved with a Gaussian kernel of \CHECK{0\farcm88}.
}
\end{figure*}

After applying the first three selection criteria described in \secref{detect}, we are left with a list of \CHECK{28} candidates in the PS1 DR1 footprint. 
We visually inspected each of these candidates to determine whether they were viable new satellite galaxies.
We find that \CHECK{most} of the remaining candidates could be clearly identified as artifacts in the PS1 DR1 catalogs.
The most obvious artifact manifested as rectangular regions of increased or decreased object density (often affecting the density of both stars and galaxies) with characteristic sizes of $\roughly 0.4 \times 0.4 \deg$ (\figref{skycell}).
These regions correspond to the \code{skycells} over which the PS1 DR1 \code{stack} images were created \citep{Flewelling:2016}.
We visually inspected these \code{skycells} and found that they contained bright stars that were causing issues in the automated image processing and catalog creation.
The corresponding under/overdensities in catalog objects bias estimates of the field density and lead to spurious candidate detection.
We masked regions of radius 0.5 deg around each of the seven \code{skycells} identified in this manner.
In addition, we find eight cases where reflected and scattered light from bright stars leads to spurious overdensities of blue objects.
These artifacts are less extreme than the \code{skycell} failures described previously, and we mask a circular region with a radius matched to the diagonal dimension of a \code{skycell} ($\roughly 0.3 \deg$).
We visually inspected the remaining \CHECK{13} candidates and found that all but one of them lacked a well-defined stellar sequence in color--magnitude space and/or a well-defined spatial morphology. 
These regions likely result from a combination of less obvious survey artifacts (\ie, mis-estimation of the sky background, excess sensor noise, or low level scattered light) and contamination from background galaxy clusters.
We mask regions of $\roughly 0.3 \deg$ around each of these seeds.
These masks were applied to our candidate list and were included in the geometric masking cuts described in \secref{detect}.

\begin{figure*}[t]
\centering
\includegraphics[width=0.8\textwidth]{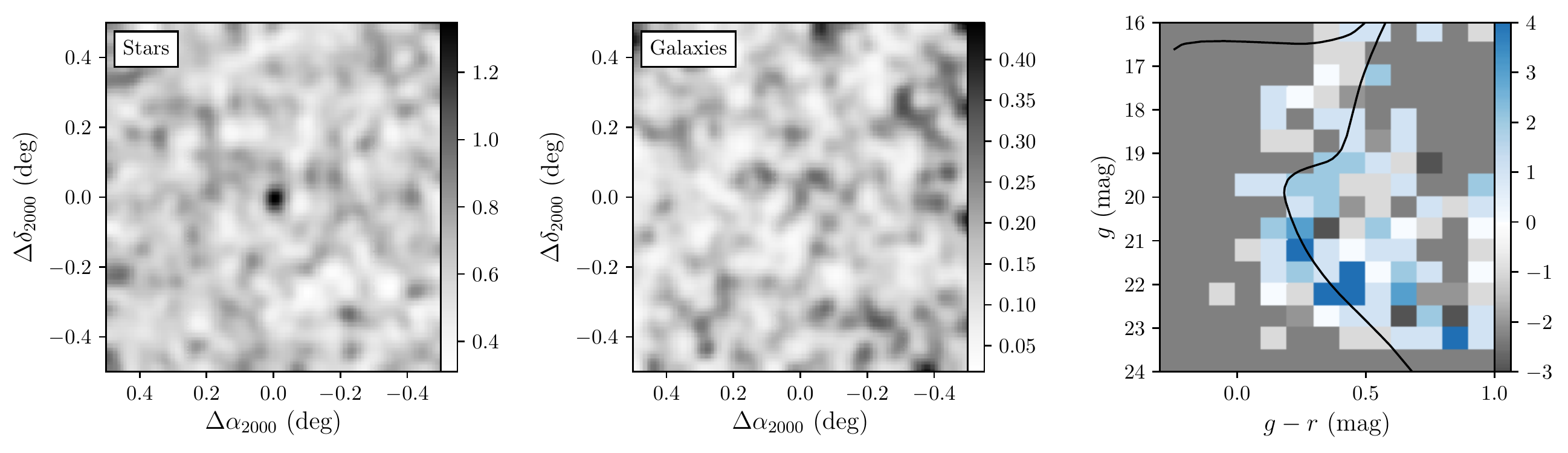}
\caption{\label{fig:cluster} 
DELVE 1 is a candidate faint outer-halo star cluster discovered in our search of PS1 DR1 data and independently in DELVE \citep{Mau:2019b}.
The small physical size ($r_{1/2} = 3.7 \pc$) and low luminosity ($M_V = 0.6$) make this system similar to other outer-halo star systems that have been discovered in recent surveys \citep[\eg,][]{Torrealba:2019,Mau:2019}.
(Left) Stellar density convolved with a Gaussian kernel of \CHECK{0\farcm67}.
(Center) Galaxy density convolved with a Gaussian kernel of \CHECK{0\farcm67}.
(Right) Hess diagram corresponding to foreground stars within $3r_h$ of the centroid and background stars in a concentric annulus from $5r_h$ to $5.8r_h$. The best-fit isochrone is shown in black.
}
\end{figure*}

As discussed in \secref{results}, one candidate passed all of our detection criteria (\figref{cluster}).
Visual inspection of this candidate did not identify any obvious survey artifacts that would significantly alter the stellar density in this region.
Similarly, we found no evidence of a correlated overdensity in the galaxy sample.
This candidate appeared similar to other recently discovered compact outer-halo star clusters \citep[\eg,][]{Torrealba:2019,Mau:2019}, and we followed the procedure of \citet{Bechtol:2015} and \citet{Drlica-Wagner:2015} to derive the physical parameters of this candidate using the maximum-likelihood fitting formalism of \ugali.
We simultaneously fit the richness, centroid position, angular extension, ellipticity, position angle, and distance modulus of this system using an affine invariant Markov Chain Monte Carlo (MCMC) ensemble sampler \citep{2013PASP..125..306F}.
Best-fit parameters and uncertainties were derived from the marginalized posterior distribution (sampled with $2.5\times10^5$ steps) and are reported in \tabref{cluster}.
During this sampling, the age and metallicity were fixed at $\age = 10\Gyr$ and $\metal = 0.0001$, respectively.
We conclude that this candidate is likely a compact ($r_{1/2} = 3.7\pc$), low-luminosity ($M_V = 0.6$) star cluster residing at a heliocentric distance of $D = 15.6\kpc$.
We note that this system was independently discovered in deeper data from the DECam Local Volume Exploration survey (DELVE), where it was investigated in more detail and named DELVE 1 \citep{Mau:2019b}.
We use the same name here to avoid confusion.

\begin{deluxetable}{c c c}
\tablecolumns{3}
\tabletypesize{\footnotesize}
\tablecaption{\label{tab:cluster}
Observed and derived properties of DELVE 1.
}
\tablehead{
\colhead{Parameter} & \colhead{Value} & \colhead{Unit}
}
\startdata
\ra & $247.725^{ +0.004 }_{ -0.004 }$ & degree \\
\dec & $-0.971^{ +0.005 }_{ -0.004 }$ & degree \\
$a_h$ & $0.8^{ +0.5 }_{ -0.3 }$ & arcmin \\
$r_h$ & $0.8^{ +0.4 }_{ -0.3 }$ & arcmin \\
$r_{1/2}$ & $3.7^{ +2 }_{ -1 }$ & pc \\
\ellip & $0.04$ & \ldots \\
\PA & $25^{ +71 }_{ -62 }$ & degree \\
\modulus\tablenotemark{a} & $16.0^{ +0.4 }_{ -0.2 } \pm 0.1$ & \ldots \\
$D_{\odot}$ & $15.6^{ +3.5 }_{ -1.1 }$ & kpc \\
$\sum p_i$ & $29^{ +5 }_{ -5 }$ & \ldots \\
\TS & $69$ & \ldots \\[-0.5em]
\multicolumn{3}{c}{ \hrulefill } \\
$M_V$\tablenotemark{b} & $0.6^{ +0.7 }_{ -1.9 }$ & mag \\
$M_{\star}$ & $86^{ +40 }_{ -31 }$ & $\Msun$ \\
$\mu$ & $26.4$ & mag arcsec${}^{-2}$ \\[+0.5em]
\enddata
\tablecomments{Uncertainties were derived from the highest density interval containing the peak and 68\% of the marginalized posterior distribution.}
\tablenotetext{a}{We assume a systematic uncertainty of $\pm0.1$ associated with isochrone modeling.}
\tablenotetext{b}{The uncertainty in $M_V$ was calculated following \citet{2008ApJ...684.1075M} and does not include uncertainty in the distance.}
\vspace{-3em}
\end{deluxetable}

\section{machine-learning classifier Parameters}
\label{app:classifier}

We modeled the observational selection function with a gradient-boosted decision tree classifier, \code{XGBClassifier}, as implemented in Python in the \code{xgboost} package version 0.82 \citep{Chen:2016}.\footnote{\url{https://xgboost.readthedocs.io}}
We trained separate classifiers on the DES and PS1 simulations.
Hyperparameters were selected using \code{GridSearchCV} from \code{scikit-learn} version 0.19.1 \citep{scikit-learn}. 
The hyperparameters scanned are $\code{learning\_rate} = \{0.01, 0.05, 0.1\}$, $\code{max\_depth} = \{6, 7, 8\}$, and $\code{n\_estimators} = \{100, 250, 500\}$, with fixed $\code{max\_delta\_step} = 1$. 
The optimal training hyperparameters for the DES simulations were $\code{learning\_rate} = 0.01$, $\code{max\_depth} = 8$, and $\code{n\_estimators} = 500$.
The optimal training hyperparameters for the PS1 simulations were $\code{learning\_rate} = 0.05$, $\code{max\_depth} = 7$, and $\code{n\_estimators} = 250$.
The trained classifiers are available at \href{https://github.com/des-science/mw-sats}{https://github.com/des-science/mw-sats}.

\end{document}